\documentclass[fleqn,usenatbib,useAMS]{mnras}

\usepackage[T1]{fontenc}
\usepackage{ae,aecompl}

\DeclareRobustCommand{\VAN}[3]{#2}
\let\VANthebibliography\thebibliography
\def\thebibliography{\DeclareRobustCommand{\VAN}[3]{##3}\VANthebibliography}

\usepackage{graphicx}	
\usepackage{amsmath}	
\usepackage{amssymb}	
\usepackage{float}
\usepackage{newtxtext,newtxmath}

\usepackage[english]{babel}

\usepackage{amsmath}
\usepackage{amssymb}
\usepackage{graphicx}
\usepackage{tablefootnote}

\setcitestyle{authoryear}
\bibpunct{(}{)}{;}{a}{}{,}

\title{White dwarf binaries suggest a common envelope efficiency  $\alpha \sim 1/3$}
\author[Peter Scherbak, Jim Fuller]{Peter Scherbak,$^{1}$ Jim Fuller$^{2}$%
\\
$^{1}$California Institute of Technology, Astronomy Department, Pasadena, CA 91125, USA\\
$^{2}$TAPIR, California Institute of Technology, Pasadena, CA 91125, USA}

\date{Accepted XXX. Received YYY; in original form ZZZ}

\pubyear{2022}

\begin{document}
\label{firstpage}
\pagerange{\pageref{firstpage}--\pageref{lastpage}}
\maketitle

\begin{abstract}

Common envelope (CE) evolution, which is crucial in creating short period binaries and associated astrophysical events, can be constrained by reverse modeling of such binaries' formation histories. Through analysis of  a sample of well-constrained  white dwarf (WD) binaries with low-mass primaries (7 eclipsing double WDs, 2 non-eclipsing double WDs, 1 WD-brown dwarf), we estimate the CE energy efficiency $\alpha_{\rm{CE}}$ needed to unbind the hydrogen envelope. We use grids of He- and CO-core WD models to determine the masses and cooling ages that match each primary WD's radius and temperature. Assuming gravitational wave-driven orbital decay, we then calculate the associated ranges in post-CE orbital period.
By mapping WD models to a grid of red giant progenitor stars, we determine the total envelope binding energies and possible orbital periods at the point CE evolution is initiated, thereby constraining $\alpha_{\rm CE}$. Assuming He-core WDs with progenitors of 0.9 - 2.0 \(M_\odot\), we find $\alpha_{\rm CE} \! \sim \! 0.2-0.4$ is consistent with each system we model. Significantly higher values of $\alpha_{\rm{CE}}$ are required for higher mass progenitors and for CO-core WDs, so these scenarios are deemed unlikely. Our values are mostly consistent with previous studies of post-CE WD binaries, and they suggest a nearly constant and low envelope ejection efficiency for CE events that produce He-core WDs.

\end{abstract}


\section{Introduction}

White dwarf (WD) binaries are not only the progenitors of many interesting astrophysical events, but they are also incredibly useful astrophysical laboratories. The discovery rate of WD binaries has greatly accelerated over the past decade, primarily due to two surveys: the Extremely Low Mass (ELM) Survey (e.g., \citealt{brown:20}) and the Zwicky Transient Facility (ZTF) survey \citep{bellm_2019}. The ELM Survey has discovered $\sim$100 WD binaries, but most are non-eclipsing and have poorly constrained companions. The ZTF survey has doubled the number of eclipsing WD binaries (e.g., \citealt{2020ApJ...905...32B}) whose masses and radii are usually well constrained. Backwards modeling from the observed properties of these binaries can lead to new insights on their formation history, especially their common-envelope (CE) phase of evolution.

The CE is a stage of binary stellar evolution where two stars orbit inside a shared envelope \citep{1976IAUS...73...75P}. CE evolution is thought to be crucial for the creation of exotic systems such as double WD binaries (DWDs) and merging neutron stars.
These and other short-period binaries that contain at least one compact object likely underwent CE evolution in the past \citep{ivanova_common_2013}. The CE is the most important phase of such a binary’s formation history, because it is responsible for the expulsion of the hydrogen (H) envelope and a drastic reduction in separation, transforming a wide binary into a short-period, compact binary.

Despite its importance, the CE is arguably the least understood stage of binary evolution, representing a limitation of binary population synthesis models. For example, changing the parametrization of the CE event can affect the theoretical rates and delay times of black hole mergers, which is important
to interpret the gravitational wave events detected by LIGO-VIRGO (e.g., \citealt{ligo_scientific_collaboration_and_virgo_collaboration_gwtc-1_2019}). The rate of Type Ia supernova, as well as the nature of their main formation channel, also depends on the parametrization of the CE \citep{ivanova_common_2013}. Therefore, uncertainty regarding the CE event propagates to different areas of astronomy. 

The most common parametrization of the CE event is based on an energy conservation approach, using an $\alpha_{\rm CE}$ parameter that is defined as the fraction of the change in orbital energy (as the inspiral occurs) that goes into the unbinding of the CE \citep{livio_common_1988}. 
Hydrodynamical simulations of the CE have produced a range of results for $\alpha_{\rm CE}$, possibly due to a failure of simulations to converge \citep{iaconi_effect_2017}. They also struggle to track the long-term (thermal time-scale) evolution, which may continue to affect the orbit. Therefore, an independent approach to estimate $\alpha_{\rm CE}$, based on observations rather than pure simulations, is valuable.

The short orbital period of many WD binaries implies that they were created through CE evolution, and they provide a unique opportunity to constrain the physics of the CE. In addition to the system's orbital period, properties such as component masses, radii, and temperatures are often reported for both WDs, using methods such as light curve fitting and radial velocity measurements. These well-constrained systems can yield detailed insights into the histories of WD binaries.

Estimates of $\alpha_{\rm CE}$ have already been reported from discoveries of WD-M dwarf binaries \citep{zorotovic_post-common-envelope_2010,de_marco__2011,davis_is_2012}, but the evolution of those systems is complicated by poorly understood magnetic braking that dominates orbital decay. Many of these systems are young and hence the braking prescription does not strongly affect their inferred post-CE orbital period \citep{schreiber_2003}, but it is still necessary to compare to other types of post-CE systems. WD binaries are simpler to model because their orbit only decays through gravitational wave (GW) emission. Hence, they can provide both comparisons to existing values of $\alpha_{\rm CE}$ and also additional constraints. For instance, if $\alpha_{\rm CE}$ values from WD binaries show agreement with $\alpha_{\rm CE}$ values from WD-M dwarf binaries, it would become more justified to extrapolate that value to all binaries formed through CE evolution and make more robust predictions for population synthesis. The growing number of well-constrained WD binaries makes it possible to report $\alpha_{\rm CE}$ for many such systems, not just a few case studies.

Either one or two phases of CE evolution may be responsible for the formation of DWD binaries. In this paper, we concentrate on the most recent phase of CE evolution, assumed to be responsible for the formation of the hotter, primary WD and also likely responsible for the most drastic reduction in separation. We are motivated by \cite{woods_formation_2011}, which found that a \textit{stable} first stage of mass transfer can explain observed systems, but that the second stage (what we model) is indeed unstable and involves a CE.

By backwards modeling of the WD primary, we can therefore constrain $\alpha_{\rm CE}$ for the final phase of CE evolution. Specifically, if the system's "birth period" can be determined - its orbital period immediately following CE ejection, when the primary WD is formed - then we can calculate the post-CE orbital energy. Combining that with the possible binding energy of progenitor stars, plus the pre-CE orbital energy, determines the energy budget for the CE.
\cite{nelemans_reconstructing_2000} and \cite{nelemans_reconstructing_2005} used similar techniques for previously discovered WD binaries, but here we additionally model more recently discovered systems from ELM and ZTF.



 

In Section \ref{HeWDs} we discuss the creation of a grid of helium (He) WDs in order to estimate cooling ages and birth periods for 10 binaries. In a similar analysis in Section \ref{COWD}, we consider the possibility of the primary WD instead being a carbon-oxygen (CO) WD. In Section \ref{progenitor main}, we construct a grid of red giant stars to model the progenitors of the binaries and determine their envelope binding energies. Section \ref{results main} presents the ranges for birth period and $\alpha_{\rm CE}$ for all systems. In Section \ref{discussion} we discuss additional uncertainties and compare our results to estimates of $\alpha_{\rm CE}$ in other works. We conclude and summarize in Section \ref{conclusion}.

\section{He WD Simulations} 

\label{HeWDs}

\subsection{Binary Systems}

\label{systems summary}

We modeled short-period WD binaries that 1) have primary masses between $\sim 0.3 - 0.45$ $M_\odot$ and are likely He-core WDs, 2) have a tight constraint on the secondary mass $M_2$, and 3) are at orbital periods greater than 20 minutes. We did not model binaries at periods lower than 20 minutes because of the associated uncertainties with strong tidal effects that could change the cooling behavior \citep{fuller_dynamical_2013}. Table \ref{tab:binary parameters} summarizes the systems we model. All except for two (WD 0957-666 and WD 1101+364) are eclipsing. Each of these systems has a primary mass $M_1 \gtrsim 0.3 \, M_\odot$ and is likely to be formed by a CE event rather than stable mass transfer \citep{li_formation_2019}.

\renewcommand{\arraystretch}{1.5}
\begin{center}
\begin{table*}
\begin{tabular}{|p{2.5cm}|p{1.2cm}|p{1.2cm}|p{2cm}|p{2cm}|p{1.8cm}|p{1.3cm}|p{1cm}|p{1cm}|}     
 \hline
 System &   \(M_1\) (\(M_\odot\)) & \(M_2\) (\(M_\odot\))  & $T_{\rm{eff,1}}$ (kK) & $T_{\rm{eff,2}}$ (kK) & \(R_1\) \, ($10^{-2} \, $\(R_\odot\)) & log g & q & $P_{\rm{b}}$ (min)    \\ [0.5ex] 
 \hline\hline
 ZTF J2029$^{\textcolor{red}{1}}$ & $0.32\pm ^{0.04}_{0.04}$ & $0.3\pm ^{0.04}_{0.04}$  & $18.25\pm ^{0.25}_{0.25}$ & $15.3\pm ^{0.3}_{0.3}$ & $2.9\pm ^{0.2}_{0.3}$ & - & - & 20.868  \\ 
  ZTF J0722$^{\textcolor{red}{1}}$ & 0.38$\pm ^{0.04}_{0.04}$ & 0.33$\pm ^{0.03}_{0.03}$  & $19.9\pm ^{0.15}_{0.15}$ & $16.8\pm ^{0.15}_{0.15}$ & $2.24\pm ^{0.04}_{0.02}$ & - & - & 23.709  \\
  ZTF J1749$^{\textcolor{red}{1}}$ & 0.40$\pm ^{0.07}_{0.05}$ & 0.28$\pm ^{0.05}_{0.04}$  & $20.4\pm ^{0.2}_{0.2}$ & $12.0\pm ^{0.6}_{0.6}$ & $2.2\pm ^{0.3}_{0.4}$ & - & - & 26.434  \\

 SDSS J0822$^{\textcolor{red}{2,3}}$ & 0.304$\pm ^{0.014}_{0.014}$ & 0.524$\pm ^{0.05}_{0.05}$  & $13.92\pm ^{0.255}_{0.255}$ & - & $3.1\pm ^{0.6}_{0.6}$ & $^*7.14\pm^{0.05}_{0.05}$ & - & 40.501  \\

 ZTF J1901$^{\textcolor{red}{1}}$ & 0.36$\pm ^{0.04}_{0.04}$ & 0.36$\pm ^{0.05}_{0.05}$  & $26.0\pm ^{0.2}_{0.2}$ & $16.5\pm ^{2.0}_{2.0}$ & $2.9\pm ^{0.1}_{0.2}$ & - & - & 40.602  \\
 
 SDSS J1205$^{\textcolor{red}{4}}$ & 0.39$\pm ^{0.02}_{0.02}$ & 0.049$\pm ^{0.006}_{0.006}$  & $23.68\pm ^{0.43}_{0.43}$ & - & $2.2\pm ^{0.03}_{0.03}$ & $^*7.37\pm ^{0.05}_{0.05}$ & - & 71.230  \\
 WD 0957$^{\textcolor{red}{5,6,7}}$ & 0.37 & - & $30\pm ^{0.5}_{0.5}$ & 11 & -  &  $^*7.285\pm ^{0.082}_{0.082}$ & $1.13\pm ^{0.02}_{0.02}$ & 87.83      \\  
   SDSS J1152$^{\textcolor{red}{8}}$ & 0.362$\pm ^{0.014}_{0.014}$ & 0.325$\pm ^{0.013}_{0.013}$  & $20.8\pm ^{1.06}_{0.96}$\rm{(eclipse)} \newline $21.2\pm ^{1.2}_{1.1}$ \rm{(SED)} & $10.4\pm ^{0.4}_{0.34}$\rm{(eclipse)} \newline $11.1\pm ^{0.95}_{0.77}$\rm{(SED)}  & $2.12\pm ^{0.03}_{0.03}$ &  $7.344\pm ^{0.014}_{0.014}$ & - & 143.806  \\
   CSS 41177$^{\textcolor{red}{9,10}}$ & 0.378$\pm ^{0.023}_{0.023}$ & 0.316$\pm ^{0.011}_{0.011}$  & $22.439\pm ^{0.059}_{0.059}$ & $10.876\pm ^{0.032}_{0.032}$ & $2.224\pm ^{0.041}_{0.041}$ & $7.322\pm ^{0.015}_{0.015}$  & - & 167.062  \\

 WD 1101$^{\textcolor{red}{8,11,12}}$ & 0.29 & - & $15.5\pm ^{0.5}_{0.5}$ & 12 & - &  $^*7.38\pm ^{0.049}_{0.049}$ & $0.87\pm ^{0.03}_{0.03}$ & 208.4 \\

 \hline

\end{tabular}

 \caption{Measured parameters for all the systems we model, including the masses of the two components, \(M_1\) and \(M_2\), the radii of the components, \(R_1\) and \(R_2\), the effective surface temperature of the components, $T_{\rm{eff,1}}$ and $T_{\rm{eff,2}}$, the surface gravity log g, the mass ratio $q \equiv \frac{M_1}{M_2}$ (if measured), and the orbital period, $P_{\rm{b}}$. An asterisk next to the log g means there is only a single value reported; otherwise, log $g_2$ is reported as well. \(M_1\) is defined as the primary via observation of the primary eclipse, except for WD 0957 and WD 1101 where it is the brighter WD. As expected, \(M_1\) is hotter than \(M_2\) when companion temperatures are reported. All companions \(M_2\) are believed to be WDs except for SDSS J1205, where the companion is a brown dwarf. Parameters are from \protect \cite{2020ApJ...905...32B}$^{\textcolor{red}{1}}$, \protect\cite{brown_discovery_2017}$^{\textcolor{red}{2}}$,
   \protect\cite{kosakowski:20}$^{\textcolor{red}{3}}$,
   \protect\cite{parsons_two_2017}$^{\textcolor{red}{4}}$,
   \protect\cite{maxted_mass_2002}$^{\textcolor{red}{5}}$,
   \protect\cite{moran_detached_1997}$^{\textcolor{red}{6}}$,
   \protect\cite{bragaglia_temperatures_1995}$^{\textcolor{red}{7}}$,
   \protect\cite{parsons_pulsating_2020}$^{\textcolor{red}{8}}$, 
   \protect \cite{bours_precise_2014}$^{\textcolor{red}{9}}$
   ,  \protect \cite{bours_hstcos_2015}$^{\textcolor{red}{10}}$,
   \protect \cite{marsh_discovery_1995}$^{\textcolor{red}{11}}$,
   \protect \cite{bergeron_spectroscopic_1992}$^{\textcolor{red}{12}}$.}
\label{tab:binary parameters}
\end{table*}
\end{center}

\vspace*{-23pt}

From a sample of WD binaries discovered with ZTF, we model ZTF J2029+1534, ZTF J0722-1839, ZTF J1749+0924, and ZTF J1901+5309  \citep{2020ApJ...905...32B}. \cite{2020ApJ...905...32B} determined radii via light curve modeling and surface temperatures via spectroscopy, then combined radius and temperature measurements to determine masses via mass-radius relations. 

We also model SDSS J082239.546+304857.19, found by the ELM Survey \citep{brown_discovery_2017}. Radii were reported via light curve modeling, temperature and surface gravity (log g) of the primary from spectroscopy, and mass from evolutionary curves fitted to temperature and log g. A later paper updated the radius of the primary and the orbital period of the system \citep{kosakowski:20}, which we use in our analysis, but did not contain an updated mass estimate of the primary. Most other binaries in the ELM Survey are not eclipsing and do not have well-constrained secondary masses, which would lead to an inherent uncertainty in the orbital evolution of the system. \cite{hermes_radius_2014} estimated $M_1$ and $M_2$ for 8 ELM WDs using the observed ellipsoidal variations - however, $M_1$ for these systems ($M_1 \lesssim 0.2 \, M_\odot$) is less than that of the other systems we model, and it is unclear whether these low-mass WDs were created through unstable or stable mass transfer.


For the WD binary CSS 41177 (SDSS J100559.10+224932.3), the mass and radius are measured from light curve plus radial velocity modeling, independent of mass-radius relations \citep{bours_precise_2014}. The temperature of the primary is again from spectroscopy, and log g values are also reported \citep{bours_hstcos_2015}.
Similarly, for the WD binary SDSS J115219.99+024814.4, the masses and radii are measured from light curve plus radial velocity modeling \citep{parsons_pulsating_2020}. Temperatures and log g values are reported both from light curve modeling and from spectroscopy. The original discovery paper \citep{hallakoun_sdss_2016}  had different results (including higher temperatures and mass estimates), but \cite{parsons_pulsating_2020} uses Gaia data to refine the measurements,
and we therefore use the more recent values.
We also model the eclipsing WD-brown dwarf (BD) binary SDSS J120515.80-024222.6 \citep{parsons_two_2017}. The masses and radii are again from light curve plus radial velocity modeling, supplemented by temperature and surface gravity measured via spectroscopy.

Finally, we model two non-eclipsing WD binaries, WD 0957-666 (2MASS J09585493-6653102) and WD 1101+364 (SDSS J110432.56+361049.0). A surface gravity value for WD 0957-666 (from the combined spectrum) is given in \cite{bragaglia_temperatures_1995},
and their reported external uncertainty is used for our modeling (see also \citealt{moran_detached_1997}).
A surface gravity value for WD 1101+364 (from the combined spectrum) is reported in \cite{bergeron_spectroscopic_1992}, and we use the average value of their reported external uncertainty in our analysis.
The mass ratio $q \equiv \frac{M_1}{M_2}$ of these double-lined systems is measured for WD 0957-666 in \cite{maxted_mass_2002} and WD 1101+364 in \cite{marsh_discovery_1995}.
\cite{maxted_mass_2002} also performed spectroscopic modeling to estimate the surface temperatures  (with uncertainties estimated to be at least 500 K) and masses for both WDs in both systems.

WD 0957-666 and WD 1101+364 are also modeled in \cite{nelemans_reconstructing_2005}, but we do not model the other binaries from that work because they either have an unconstrained companion mass or the mass of the primary is $\gtrsim 0.45$ $M_\odot$ and more consistent with a CO WD as opposed to a He WD. In particular, we do not model WD 1704+481 because \cite{maxted_mass_2002} found that the He WD that \cite{nelemans_reconstructing_2005} assumed to be the primary (the more recently formed WD) likely formed second. Instead, the more massive CO WD in the binary likely formed after the most recent stage of mass transfer.

For all systems, the primaries have a hydrogen atmosphere. Several of the binaries (those with measured values of both $T_{\rm{eff,1}}$ and $T_{\rm{eff,2}}$) are double-lined DA binaries. The two without reported $T_{\rm{eff,2}}$ are single-lined, with hydrogen absorption lines only detected from the primary.

\subsection{Creation of He WD Models} \label{create_He}

All models/simulations are performed using the MESA stellar evolutionary code \citep{Paxton2011, Paxton2013, Paxton2015, Paxton2018, Paxton2019}. We used MESA version 12778, except for cooling WD models that included H flashes and Roche-lobe overflow (RLOF) in Sec. \ref{flashing main}, for which we used version 10398.  MESA inlists used in this work are provided in a Zenodo repository \citep{peter_scherbak_2022_7272761}.

WD models were created by evolving a star up the red giant branch (RGB) until its He core reached a desired mass, then stripping away most of its envelope to leave the core as a WD, as briefly outlined in \cite{Paxton2018} and also \cite{burdge_orbital_2019}. Unless noted, we we define the boundary of the He core to be where $X_{\rm{H}}$ < 0.01 and $X_{\rm{He}}$ > 0.1 (default in MESA). As an example, to create a 0.32 \(M_\odot\)  WD with a 0.318 \(M_\odot\) He core, our specific steps are:

    \textbf{1:} Create a star via \textit{create\_pre\_main\_sequence\_model = true} with some \textit{initial\_mass} and \textit{initial\_z}. We use Z = 0.02 for our main grid, but we investigate the effect of changing metallicity on WD models in Appendix \ref{metallicityHe}. Different progenitor masses can have an impact on the ultimate behavior of the WD models (Appendix \ref{He progen}), but we found the cooling behavior similar for initial masses of 0.9 - 2.0 $M_\odot$ . This covers the most likely CE progenitor mass for He WDs, which is roughly between 1 and 1.3 $M_\odot$ \citep{li_formation_2019}. For 2 - 3 $M_\odot$ progenitors (a less likely channel), the newly-formed core is less degenerate, and can cool differently. Above about 3 $M_\odot$, we do not expect a RGB star to be able to form a He WD through the CE channel (see Sec. \ref{progenitormass}).  We therefore used a 1.2 $M_\odot$ progenitor to create the main grid of models.

   \textbf{ 2:} Evolve star on the main sequence and up the RGB until its He core reaches 0.318 $M_\odot$ via \textit{he\_core\_mass\_limit}.
    
    \textbf{3:} Use MESA's method \textit{relax\_mass} to strip away most of the star's H-rich envelope, until \textit{new\_mass} = 0.32 $M_\odot$. This leaves only a small H envelope atop the He core.
    
    \textbf{4:} Allow the newly-formed WD to adjust to the mass loss. In all cases, the model experiences an increase in surface temperature and radius, with the surface temperature increasing from $\approx$10,000 K immediately after relax mass, to 40,000 - 100,000 K depending on the WD's mass, before beginning to cool. In most models, this adjustment period is irrelevant (duration <1,000 years). However, if there is excess H on the model, there will be elevated H-burning that can last for $> \! 10^{5}$ years and cause the envelope to temporarily inflate.
    As we expect the binaries we model to be at small separations, we reject WD models that expand beyond $1 \, R_\odot$. For a typical $M_1$ of 0.4 $M_\odot$, a typical $M_2$ of 0.3 $M_\odot$, and a maximum post-CE period of about 200 minutes, the orbital separation is $\simeq$1.0 $R_\odot$. Therefore, none of our systems can fit a WD that expands beyond this value.  This restricts the upper age of this adjustment phase at $10^{6}$ years, which is small compared to the resulting cooling ages. See Fig. \ref{fig:allowed mass} for the upper mass of H this sets in our grid.
    We save the model for continued analysis when it begins to cool, i.e. when $T_{\rm{eff}}$ begins to decrease, and define this as a cooling age of zero.

\subsection{Grid of WD models}

The model saved in the previous step, simulating a hot and newly formed WD, can now be introduced into a binary simulation. We run a grid of models with WDs of different total masses and different masses of hydrogen. The lowest measured primary mass we model is reported as 0.32$\pm ^{0.04}_{0.04}$ \(M_\odot\) and the highest as 0.4$\pm ^{0.07}_{0.05}$ \(M_\odot\), and the masses in our grid ranged from 0.24 to 0.45 \(M_\odot\) in steps of 0.01 \(M_\odot\). We found it difficult to create a He WD at 0.46 \(M_\odot\) or above, as these models would instead burn He and form CO cores. 

We include diffusion and gravitational settling for all models, but only during their cooling phase (after the post-relax mass adjustment above). These processes are important to include for two major reasons:
\begin{itemize}
    \item Turning on diffusion noticeably increases a WD model's radius as in \cite{althaus_diffusion_2000}, who demonstrated the effect both through simulations and through an analytic argument involving the effect of the partially-degenerate H envelope.
    \item Turning on diffusion changes which models undergo H-burning flashes, which drastically changes their subsequent cooling behavior. Because of the importance of the CNO cycle to H flashes (e.g. \citealt{istrate_models_2016}), we included the dominant isotopes of carbon, nitrogen and oxygen in addition to H and He, in our list of elements for diffusion. 
\end{itemize}
The most important diffusion parameter to include was \textit{diffusion\_use\_cgs\_solver = .true.}, which correctly accounts for degeneracy when performing diffusion \citep{Paxton2018}.

\subsubsection{Hydrogen bounds}

\begin{figure}
    \centering
    \includegraphics[scale=0.45]{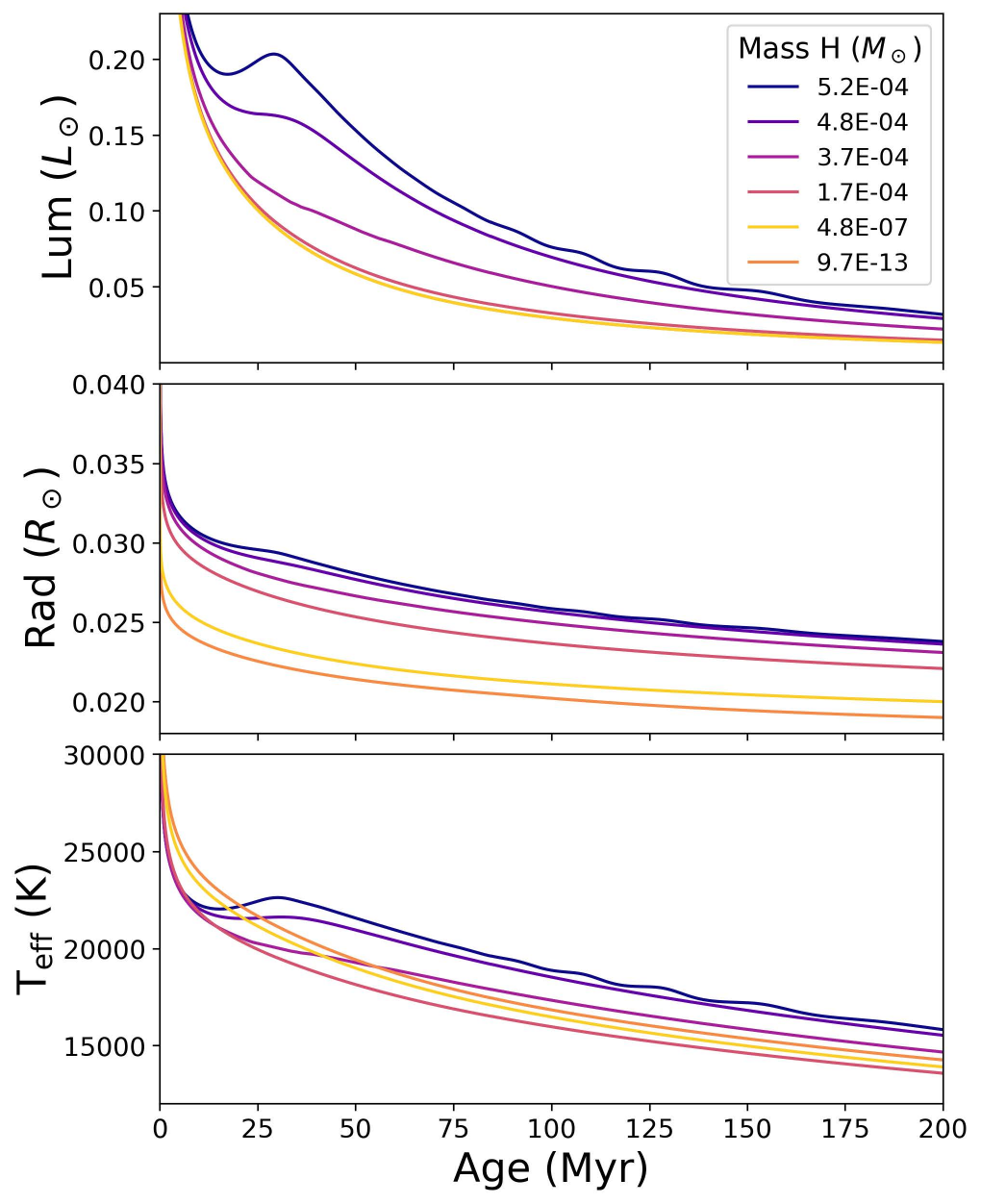}
    \caption{The effect of changing the initial hydrogen envelope mass on the cooling behavior of non-flashing WD models with a fixed total mass of $0.35 \, M_\odot$.
    As the H-mass is reduced, the WD's luminosity (top panel), radius (middle panel), and temperature (bottom panel) all decrease faster. The top (dark blue) curve is a model with residual H-burning and narrowly avoids a burning flash. Below $\approx \! 10^{-7} \, M_\odot$, further reducing H has a diminishing effect on evolution.} 
    
    
    \label{fig:coolinglumradtemp}
\end{figure}

Overall, there are 3 regimes in WD cooling behavior delineated by the amount of H on a model (See also Fig. \ref{fig:allowed mass}). For large H masses, elevated H-burning occurs and the WD expands beyond 1 $R_\odot$, and such models are not considered in our grid.

 

For moderate H masses, delayed H-burning flashes can occur (e.g., \citealt{istrate_models_2016}). In this regime,
unstable H-burning flashes lead to increases in surface temperature and expansions in radius, likely leading to RLOF with the companion. Such simulations are more complicated to run, and are discussed in Sec. \ref{flashes}.
For low H masses, the models steadily cool without ever undergoing a H flash. Both flashing and non-flashing models models may have an early stage of elevated H-burning post-relax mass, and begin to cool once the mass of H has dropped low enough.

For non-flashing models of a given mass, a varying mass of H leads to a range of cooling behavior (Fig. \ref{fig:coolinglumradtemp}). In general, the trend with decreasing H mass is for the model to both contract and cool faster.
A model that nearly flashes (upper curve, with $ \approx \! 5.2 \times 10^{-4} \, M_\odot$ H) has a brief period of elevated luminosity, temperature and radius. In addition, at high H masses, small changes in H mass (i.e. from $\approx \! 5.2 \times 10^{-4} \, M_\odot$ H to $\approx \! 4.8 \times 10^{-4} \, M_\odot$ H) have a significant affect on cooling behavior. Therefore, it is important to have densely sampled high-H models in our grid. At smaller H masses, the effect of changing H is less pronounced - between the curves representing $\approx \! 10^{-7} \, M_\odot$ H and $\approx \! 10^{-12} \, M_\odot$ H there is only a relatively small decrease in radius and almost no change in temperature and luminosity. Note in the case of Fig. \ref{fig:coolinglumradtemp} we show only a subset of our grid of models of varying H mass. 

The lower bound of H mass used in our grid is constrained by the classification of the WD primaries we model as DA WDs, i.e., the presence of H lines in their atmosphere. Therefore, we do not want to completely remove hydrogen from our WD models. We found that when the mass of H was reduced below about $10^{-8} \, M_\odot$, H would cease to dominate the model atmosphere. Even with element diffusion/gravitational settling turned on, the fraction of H in the model's outer cells would never rise above about 0.01. Therefore, the lower mass of H included in our grid is $10^{-8} \, M_\odot$. We show a model with   $\approx \! 10^{-12} \, M_\odot$ H in Fig. \ref{fig:coolinglumradtemp} to demonstrate that, even if such a model was included, it would make little difference in cooling behavior. Including such models \textit{would}, however, slightly change the modeled mass ranges for some of our systems (e.g. a 0.25  $M_\odot$ WD is only a good fit for ZTF J2029 when having extremely low H masses of $\approx \! 10^{-12} \, M_\odot$). The modelled lower mass bound would change by 0.01 $M_\odot$ in such cases.



\subsection{Non-flashing models in a Binary:  Determining Birth Period}

\label{binary cool}

\begin{figure}
    \centering
    \includegraphics[scale=0.52]{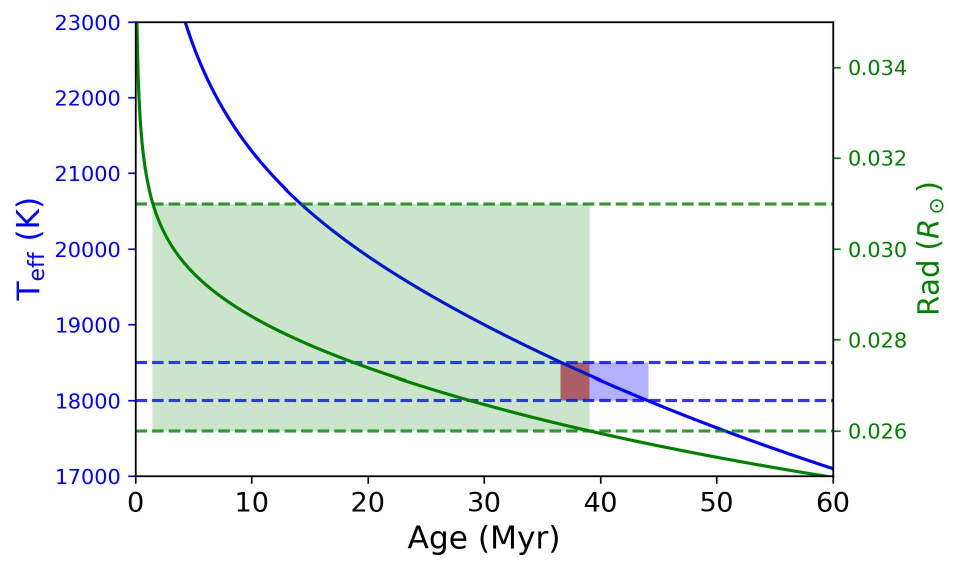}
    \caption{In blue (green): temperature (radius) curve of a cooling 0.34 $M_\odot$ WD. Dashed lines show temperature/radius constraints of ZTF J2029's primary. Blue/green shaded regions demonstrate allowed age and temperature/radius ranges independently. Their intersection in red gives the model's age range consistent with both temperature and radius. Repeating this exercise for all primary models (of varying total mass and H mass) traces out the allowed age range.}
    \label{fig:cooling}
\end{figure}

In cases where H-burning flashes do not occur, the only evolution of the binary system is for the WDs to cool and inspiral toward one another due to the emission of GWs. We assume no magnetic braking for the low-mass WDs we model. For all systems (except SDSS J0822, WD 0957-666 and WD 1101+364, see \ref{SDSS J0822} and \ref{Nelemans WDs sec}), we determine which WD models are good fits by finding the age intervals at which both their radius and temperature match observational constraints for each primary (see Fig. \ref{fig:cooling} for example). The end result for each binary (more accurately, each primary) is a collection of ``matching'' WD models with an associated mass and cooling age (see Table \ref{tab:matching_models}).
Accounting for any initial phase of elevated H burning would only affect such ages by less than 1 Myr (Sec. \ref{create_He}), which is generally quite small compared to the possible cooling age range.

Given a matching model with some $M_1$ and some age, we use the following method to find the associated birth period following the end of the CE. There is an additional intrinsic uncertainty associated with $M_2$, for which we use the values in the literature (i.e. the values in Table \ref{tab:binary parameters}, going to $\pm 1 \sigma$).

\begin{itemize}
    \item Via Kepler's Third Law, convert the binary's observed period to a separation.
    \item With $M =M_1 + M_2$, integrate \begin{equation} \label{eq:inspiral}
    \frac{da}{dt} = \frac{-64 G^3 M_1 M_2 M}{5 a^3 c^5}  
\end{equation}
    to find the birth separation of the binary \citep{peters_gravitational_1964}. This assumes that gravitational radiation is solely responsible for the inspiral, and is valid for circular orbits, which is a reasonable assumption since systems are expected to
    leave the CE with an eccentricity $\lesssim 0.1$. \citep{ivanova_common_2013}
    \item Convert the birth separation to the birth period.
\end{itemize}
With this approach, age (determined by the behavior of the temperature/radius cooling curves of the primary) is the fundamental property of a matching model, and acts as a proxy for the birth period of the binary. For all non-flashing models, we verified that the radius remained well below the Roche lobe so that our assumption of mass transfer not occurring is valid.


In general, the matching model with the maximum age has a relatively high mass and a high mass of H. In contrast, the matching model with the minimum age has a relatively low mass and a low mass of H. As expected, WDs with higher masses tend to cool slower, given their reduced radius. In addition, the presence of a significant H envelope helps prevent cooling. See Fig. \ref{fig:flash_vs_noflash} for an example of the slowest and fastest cooling models for SDSS J0822, also with a comparison to flashing models. The mass range of matching models is in good agreement with the quoted mass of the primaries from the literature (Table \ref{tab:matching_models}). This is expected since several of the systems had their masses inferred from theoretical mass radius relations.

For two systems (CSS 41177 and SDSS J1152), the reported mass $M_1$ is from radial velocity and light curve modeling, independent of theoretical cooling models/mass-radius relations. Therefore, we treat $M_1$ as an independent observational constraint enforced for our range of matching models.
This does not significantly change the estimated birth periods for these two systems, but it does ultimately help  restrict the range of possible $\alpha_{\rm{CE}}$ values (which depends on the modelled WD mass).

Apart from CSS 41177 and SDSS J1152, the above analysis fitted models only to observed surface temperature and radius. For four of the systems (CSS 41177, SDSS J1152, SDSS J1205, SDSS J0822), surface gravity data is also reported via spectroscopy. For the first three of these, applying that measurement (using the $\pm 1 \sigma$ range in log g) produced no additional constraint because almost all models that matched to temperature and radius were also consistent with log g. So for CSS 41177, SDSS J1152, and SDSS J1205, we do \textit{not} use the surface gravity constraints.

\subsubsection{SDSS J0822}

\label{SDSS J0822}

However, the surface gravity measurement does affect results for SDSS J0822, because of tension between surface gravity and radius constraints. The WD primary's radius is not well constrained from light curve fitting, with a value of $0.031 \pm 0.006$ $R_\odot$ \citep{kosakowski_multiband_2021}. However, \cite{kosakowski_multiband_2021} also performed an analysis assuming that the mass of the primary was $0.304 \pm 0.014 $ $M_\odot$ (which was reported in the discovery paper \citealt{brown_discovery_2017}, using $T_{\rm{eff}}$ and log g). Assuming this mass and using WD cooling tracks gave a model radius of $0.025 \pm 0.001 $ $R_\odot$, just within 1$\sigma$ of the light curve constraint.

In our analysis, when we fit to $T_{\rm{eff}}$ and radius constraints alone (the latter from light curve modeling), our modeled mass range extends from $0.22 - 0.32$ $M_\odot$, mainly because of the high radius errorbars. 
Only the primary's log g value is reported for this system, but the contribution of the unseen secondary is likely small. If we fit to $T_{\rm{eff}}$, radius, \textit{and} log g, our modeled mass is reduced to only $0.30 - 0.32 $ $M_\odot$ (overlapping well with the mass estimate from \citealt{brown_discovery_2017}). Discarding the radius and fitting to $T_{\rm{eff}}$ and log g slightly broadens the range to $0.28 - 0.32 $ $M_\odot$ (therefore the radius constraint is still useful).
Therefore, for our results, we fit to $T_{\rm{eff}}$ and log g (from \citealt{brown_discovery_2017}) and radius (from \citealt{kosakowski_multiband_2021}) to find the range of masses/cooling ages for SDSS J0822.


\subsubsection{WD 0957-666 and PG 1101+364}
\label{Nelemans WDs sec}

Unlike the systems discussed above, WD 0957-666 and PG 1101+364 have measured log g values but do \textit{not} have measured radii. So for these models, we instead fit our cooling curves to log g and surface temperature to find matching models and the associated cooling age. \cite{maxted_mass_2002} reported surface temperatures for both WDs in each system (Sec. \ref{systems summary}). However, only a combined value of log g is reported (instead of fits to both WD's atmospheres). For WD 0957-666, $T_{\rm{eff,1}}$ is much larger than $T_{\rm{eff,2}}$ (30,000 K vs 11,000 K) so the contribution of the secondary likely does not matter greatly. \cite{maxted_mass_2002} estimated a primary mass of 0.37 $M_\odot$ by accounting for both WDs and using their luminosity ratio, the same value estimated by \cite{moran_detached_1997} by simply fitting a single star model.

For WD 1101+364, the temperatures of the WDs are closer together (15,500 K vs 12,000 K) and so it is not justified to ignore the secondary. For our upper mass limit, we assume the spectrum is dominated by the primary (lower mass WD), and fit to $T_{\rm{eff, 1}}$ and the combined log g, which leads to matching models from 0.34 - 0.40 $M_\odot$ (likely too large compared to the true value). For our lower mass limit, we assume the spectrum is dominated by the secondary and fit $T_{\rm{eff, 2}}$ and the combined log g to find values for $M_2$. Then, we use the measured mass ratio $q$ to give the range of primary masses as 0.27 - 0.34 $M_\odot$. To determine the cooling age associated with this lower mass range, we fit to $T_{\rm{eff, 1}}$ alone - however, any ambiguity in cooling age/birth period is less important than the large uncertainty in mass for this system. In summary, we estimate the mass of WD 1101+364's primary to be between 0.27 and 0.4 $M_\odot$.

The other difference for these 2 systems is that the mass ratio $q$ is reported instead of $M_2$. Therefore, for each WD primary model with mass $M_1$, we calculate the associated range of $M_2$ values (going to $\pm 1 \sigma$ in q). Then, the cooling ages can be converted to a range of birth periods as before.

\renewcommand{\arraystretch}{1.5}
\begin{center}
\begin{table*}
\begin{tabular}{|p{1.8cm}|p{1.5cm}|p{1.5cm}|p{2.3cm}|p{2.3cm}|p{2.6cm}|p{2.4cm}|p{0.0cm}|}     
 \hline
 System &  $M_1$ ($M_\odot$)  & Modeled mass  ($M_\odot$) &    Cooling age (Myr) & Birth period (min)    & Mass of flashing models ($M_\odot$)  & Flashing models, birth period  \\ 
 \hline\hline

 ZTF J2029 & 0.32$\pm ^{0.04}_{0.04}$& 0.26-0.35 &2.4 - 112.5  &23.3 - 70.8   &   0.26 - 0.34    & 35 - 66 (31 - 69)    \\
 ZTF J0722 &0.38$\pm ^{0.04}_{0.04}$&0.34-0.4 &36.5 - 106.2   & 47.0 - 74.2&   0.39    & 58  - 64 (56 - 60) \\
  ZTF J1749 &0.40$\pm ^{0.07}_{0.05}$&0.32-0.45& 15.2 - 118.6  & 36.0 - 78.5    & 0.37 - 0.43 (0.33 - 0.43) & 48 - 71 (44 - 68)   \\ 
  SDSS J0822 &0.304$\pm ^{0.014}_{0.014}$& 0.3-0.32  & 93.6 - 265.0  & 75.1 - 113.5 & 0.31 - 0.32 & 83 - 104 (79 - 96)   \\  
 ZTF J1901 &0.36$\pm ^{0.04}_{0.04}$&0.3-0.39 & 0.46 - 25.0  & 40.8 - 54.7 &   0.31 - 0.39 (0.29 - 0.39) &   44 - 63 (44 - 70)     \\  
  SDSS J1205 &0.39$\pm ^{0.02}_{0.02}$&0.38-0.43 &19.8 - 60.5  & 72.0 - 74.2&  0.42 - 0.43   & 72 - 73   \\  
 WD 0957 & 0.37 & 0.38 - 0.45 & 2.0 - 22.2 & 88.2 - 92.6 & 0.41 - 0.43 (0.4 - 0.43) &  89 - 90 (88 - 90)      \\ 
 SDSS J1152 &0.362$\pm ^{0.014}_{0.014}$& 0.36 - 0.38 & 31.5 - 54.2  & 146.0 - 147.8 &  n/a   & n/a  & \\  
 CSS 41177 &0.378$\pm ^{0.023}_{0.023}$ & 0.37-0.4 & 25.5 -  31.5  & 168.4 - 168.9  &   n/a  & n/a   \\

 WD 1101 & 0.29 & 0.27 - 0.4 & 5.8 - 272.0 & 208.6 - 222.7 & 0.37 - 0.4 & 215 - 222 (215 - 221)  \\
\hline

\end{tabular}

 \caption{From left to right: System, measured mass range \(M_1\), modeled mass (without flashes), modeled cooling age (without flashes), modeled birth period (without flashes), modeled mass (with flashes), modeled birth period (with flashes). The rightmost columns assume non-mass-transferring flashing models, while numbers in parentheses account for models undergoing mass transfer at an orbital period of 100 minutes and with a companion mass of 0.3 $M_\odot$.}
\label{tab:matching_models}
\end{table*}

\end{center}

\subsection{Flashing models in  a Binary}

\label{flashing main}

\subsubsection{Basics}

%

\begin{figure}
    \centering
    \includegraphics[scale=0.6]{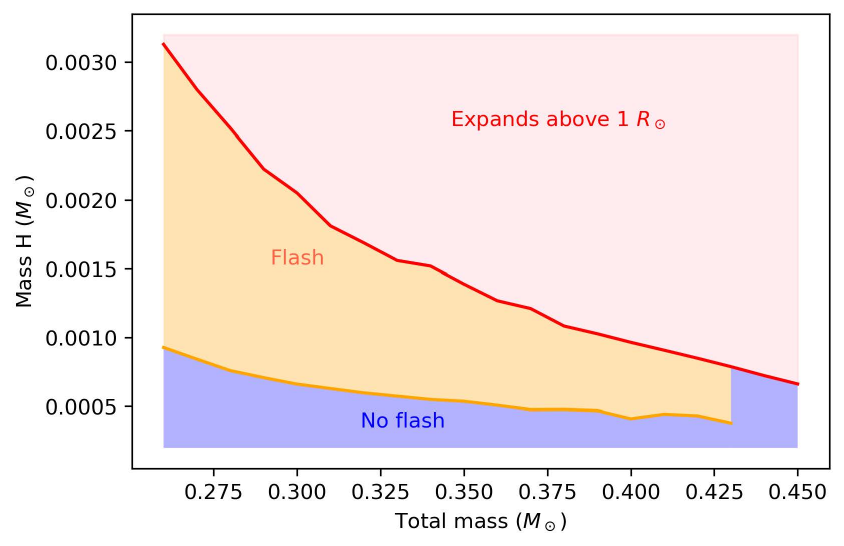}
    \caption{WD cooling behavior as a function of WD mass and initial hydrogen mass. Red region: Model has elevated H burning and expands beyond 1 $R_\odot$. The red line is the upper bound of hydrogen included in our main grid. Orange region: model will begin to cool as a proto-WD \citep{istrate_models_2016} but then undergo one or more H flashes. Blue region: model will cool without ever experiencing a flash. Above 0.43 $M_\odot$, models do not flash.
    }
    \label{fig:allowed mass}
\end{figure}

\label{flashes}

\begin{figure}
    \centering
    \includegraphics[scale=0.5]{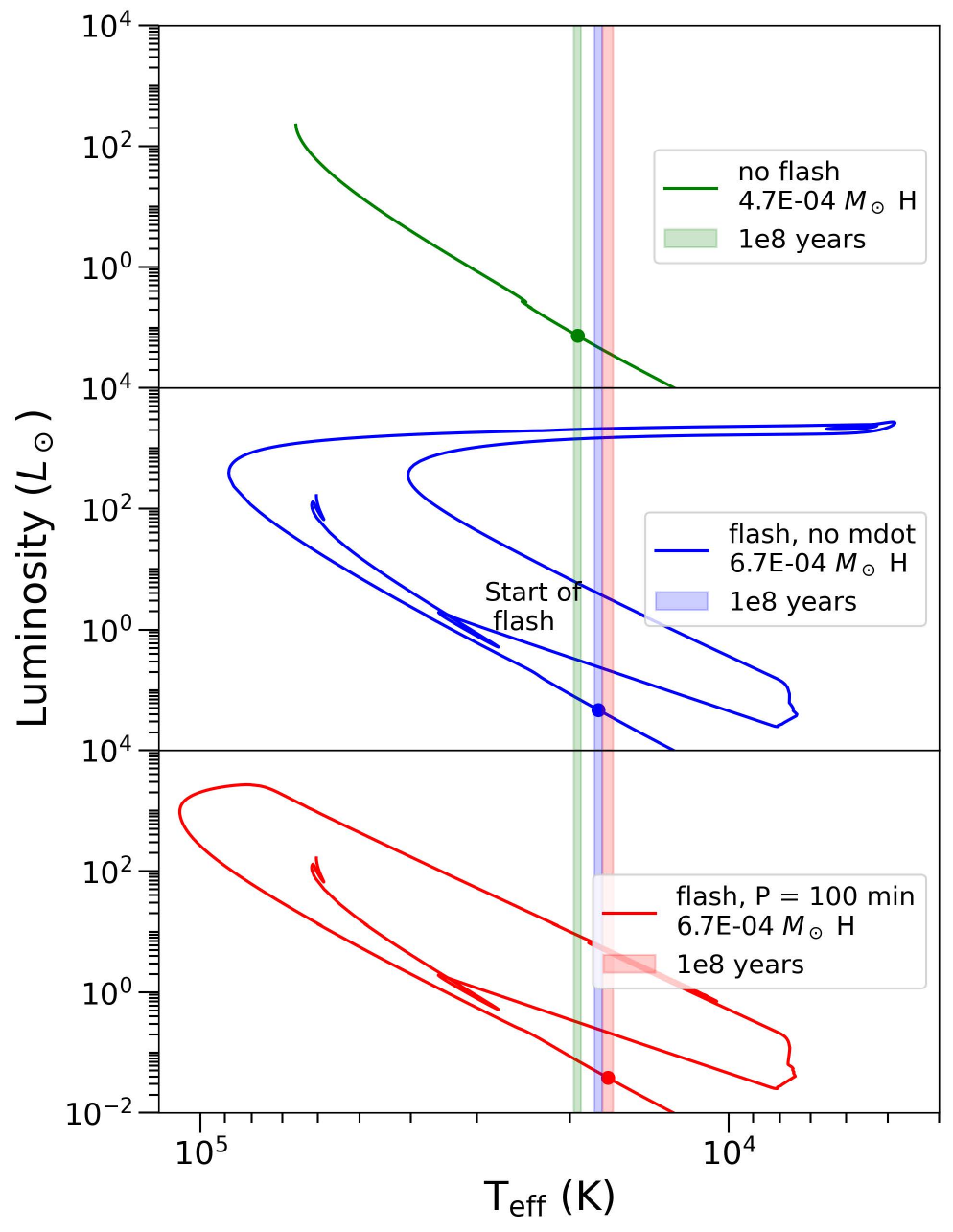}
    \caption{HR diagrams for three 0.37 $M_\odot$ models, with different initial masses for H envelopes. Upper: model undergoes leftover, stable H-burning and avoids H flash. Middle: model undergoes H flash, but mass transfer is turned off. Lower: model undergoes H flash, and remains at constant radius once RLOF begins. Coloured bars represent when the model is $\approx$ 1e8 years old. Therefore, the non-flashing model cools more slowly in the long-term than flashing models. }
    \label{fig:flash HR}
\end{figure}

A H flash occurs when
semi-degenerate H-burning ignites unstably, similar to H-novae in accreting WDs, creating enormous increases in burning luminosity \citep{althaus_new_2013}.
Flashing behavior is made more likely by diffusive mixing (e.g. \citealt{istrate_models_2016}). 
After the onset of the flash, the star expands greatly. If the model was allowed to continue without RLOF, the radius would continue to expand to tens of $R_\odot$ (middle panel of Fig. \ref{fig:flash HR}). However, in a binary simulation, the model undergoes RLOF and its radius remains fixed at its Roche-lobe value (bottom panel of Fig. \ref{fig:flash HR}).  Our MESA binary simulations included the flashing model as the primary, and a point mass as the secondary. For the bottom panel of Fig. \ref{fig:flash HR}, the simulation is performed at an orbital period of 100 minutes and with a companion of 0.3 $M_\odot$. See also Appendix \ref{appendix flashes} for discussion of the mass transfer prescription and effect of orbital period and companion mass.

Once nuclear burning has subsided, the model's envelope can begin to contract and cool once more. Overall, the flash and subsequent RLOF cause the model to do a loop on a HR diagram, making WD cooling more complicated to model. Fig. \ref{fig:allowed mass} shows which models do and do not flash (showing results only down to a mass of $0.26$ $M_\odot$, which are the masses of models relevant to our systems). 

We find that $\sim$75 per cent of the pre-flash H is either lost to mass transfer or burnt to He during a H flash. The mass transfer rate can be very large, approaching $10^{-3}$ $M_\odot/\rm{yr}$ at its greatest, and leads to a significant stripping of the H envelope, with total mass loss sometimes above $10^{-3}$ $M_\odot$.  The speed of our model runs is heavily dependant on what orbital period mass transfer occurs at, as that sets how much material is expected to overflow the Roche-lobe. Therefore, runs at shorter orbital periods are more challenging than runs at longer orbital periods.

For flashing models including mass transfer, we found the effect on the orbit to be small, and the subsequent orbital evolution is nearly identical to that which includes GW-induced inspiral alone.
The reason is that even if the entire H envelope is lost (at most $\approx 10^{-3}$  $M_\odot$), it will have little effect on the orbital separation, which to order of magnitude changes as $\Delta a/a \sim \Delta M/M$. Such small changes are much smaller than other uncertainties, so we can again safely ignore the effect of mass transfer. For our fiducial simulations, we assumed conservative mass transfer, but changing that prescription led to only small changes in the orbit or the change in the companion mass.

\subsubsection{Do flashing models provide upper/lower age bounds?}

While we have argued that mass transfer due to a H flash does not greatly affect the system's orbit, mass transfer certainly can affect the cooling of the primary. In no binary did we find the cooling curves pre-flash to be a good fit to constraints, as the radii of pre-flash models were too large.

Post-flash comparison is more difficult.  Even though a flashing model temporarily becomes hotter after a flash, it will cool faster in the long term compared to a model that is borderline to flashing, but maintains stable-burning. The stable-burning model retains a high amount of H that inhibits its long-term cooling, whereas the flashing model loses that H in its flash and cools faster.
This is demonstrated by the vertical shaded regions in Fig. \ref{fig:flash HR}, showing the location of the model on the HR diagram at a fixed age.
Therefore, non-flashing models are likely to produce longer maximum cooling ages than flashing models. 

To test this hypothesis, we ran a full grid of flashing, non-mass-transferring models as well as a grid of flashing models that undergo mass transfer at an orbital period of 100 minutes and a companion of 0.3 $M_\odot$. These grids sampled the flashing region of Fig. \ref{fig:allowed mass} (i.e. models from 0.26 to 0.43 $M_\odot$, with mass H in the orange flashing region). The results (summarized in Table \ref{tab:matching_models}) for primary masses and birth periods do not greatly differ between these two flashing grids. Therefore, modeling the mass transfer itself does not greatly affect results. This is because models will burn most of the H to He even if no H is lost due to RLOF. When comparing individual models, models with mass transfer always cooled faster post-flash. However, the upper bound on birth period is not necessarily lower when comparing the two flashing grids, because different models can be matching in different grids. 

For all but two systems, a subset of the flashing models were consistent with observed characteristics, meaning that these WDs could have undergone a flash in the past. However, for the binaries we modeled, we find that their upper and lower age bounds (as well as birth period bounds) are almost always set by non-flashing models. Note that we are \textit{not} comparing models of the same mass - just saying that when compared to the entire grid of non-flashing models, flashing models do not set the bounds.



\begin{figure}
    \centering
    \includegraphics[scale=.55]{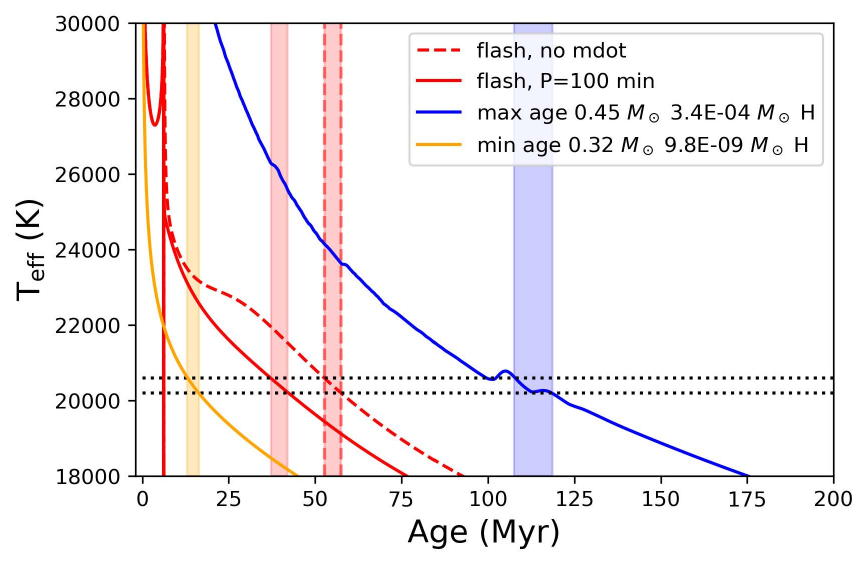}
    \caption{For ZTF J1749, we plot four models whose temperature and radius match observational constraints (radius is not shown for simplicity). The dotted lines are the observed temperature range and vertical shaded bands show the ages at which each model matches this temperature. The model with the maximum age (blue curve) is a non-flashing model that is close to flashing and slowly cools. The model with the minimum age (yellow curve) has a thin H envelope and cools quickly. Two flashing models, one with mass transfer that occurs at an orbital period of 100 min (red solid line), and one without any mass transfer (red dashed line) are also shown. Non-flashing models therefore produce the minimum and maximum possible cooling ages.}
    \label{fig:flash_vs_noflash}
\end{figure}

Fig. \ref{fig:flash_vs_noflash} summarizes the above argument, with all models shown matching observational constraints for ZTF J1749. The upper age bound comes from a non-flashing model with H mass slightly below the threshold required to flash, which cools slowly and with a tell-tale bump in temperature (where it nearly flashes) before it continues cooling. Flashing models, whether mass transferring or not, give ages below this upper bound.
While this is only one system and one set of flashing models, we could construct similar figures in almost all cases.


In one system (ZTF J1901), flashing models \textit{do} give the upper age/birth period bound, but this only occurs when the model matches observations during a very brief period of time right after a flash. This is discussed further in Appendix \ref{appendix flashes}. 

Another important result is that the grid of flashing models does not change the estimated mass range for the systems (the masses in the flashing column lie within the mass range of the non-flashing models in Table \ref{tab:matching_models}). This  means that flashing models will not significantly increase uncertainties in $\alpha_{\rm{CE}}$. Again, there is a potential exception for ZTF J1901, where the flashing grid run at a period of 100 minutes introduces a lower mass model, but it is again only a good fit for a brief period of time after the flash.


\section{CO-core WD simulations}

\label{COWD}

Although the WDs we model have $M_1 \lesssim 0.4 \, M_\odot$ and are most likely He-core WDs, it is possible that some are in fact CO-core WDs (sometimes referred to as hybrid WDs), which formed from stars with $M \gtrsim 2 M_\odot$ that ignited He-burning at smaller He-core masses than lower mass stars. Hence, we must consider this possibility in our modeling, as it entails significantly different progenitor masses and evolutionary histories.

\begin{center}
\begin{table*}
\begin{tabular}{||c c c c c||} 
 \hline
 System &  Observed (\(M_\odot\)) & Matching CO models (\(M_\odot\)) & Cooling age (Myr) & Birth period (min) \\ [0.5ex] 
\hline
 ZTF J2029 & 0.32$\pm ^{0.04}_{0.04}$  &0.32 & 998 - 1074 & 143 - 160  \\
 \hline
 ZTF J0722 & 0.38$\pm ^{0.04}_{0.04}$  &0.33 - 0.38& 436 - 707 & 115 - 143  \\
 \hline
 ZTF J1749 & 0.4$\pm ^{0.07}_{0.05}$ & 0.33 - 0.45 & 186 - 1086 & 84 -161  \\
 \hline
 ZTF J1901 & 0.36$\pm ^{0.04}_{0.04}$ & 0.33 - 0.37 & 435 - 1064 & 114-172  \\
 \hline
  SDSS J1205 & 0.39$\pm ^{0.02}_{0.02}$  & 0.37 - 0.41 & 305 - 458 & 82 - 89  \\
   \hline
WD 0957 & 0.37 & 0.35 - 0.42 & 243 - 602 & 124 - 145    \\  
\hline
 SDSS J1152 & 0.362$\pm ^{0.014}_{0.014}$  & 0.35 - 0.38 & 414 - 624 & 169 - 180  \\
  \hline
  CSS 41177 &0.378$\pm ^{0.023}_{0.023}$ & 0.35 - 0.4 & 330 - 496 & 184 - 196  \\
  \hline

 WD 1101 & 0.29 & 0.33 - 0.39 & 493 - 778 & 228 - 240  \\
 
 \hline

\end{tabular}

 \caption{From left to right: System, observed mass range \(M_1\); mass range, age range, and birth period range of our matching CO WD models. There were no matching CO WD models for the system not shown (SDSS J0822).}
\label{tab:CO result tab}
\end{table*}

\end{center}

\subsection{Creating Grid of Models}

In a similar process to the creation of He-core WD models, we create CO-core models through mass stripping of an evolved stellar progenitor. We find that progenitors from about 1 to 4  \(M_\odot\) can create CO WDs with masses that are relevant to the binaries we model (Sec. \ref{progenitormass}). The progenitor mass can potentially affect the WD cooling behavior - however, we found that such differences are generally small (see Sec. \ref{progenitor effect} for further discussion). Therefore, except where noted, we  used  a 3  \(M_\odot\) progenitor to create our grid of CO WD models.

To create a grid of WD models of a given mass (e.g., 0.4 \(M_\odot\)) our specific steps are:

\begin{itemize}
    \item Evolve a 3  \(M_\odot\) star up the RGB until its He core  reaches values close to but beneath  0.4 \(M_\odot\) (e.g., 0.39 - 0.3999 \(M_\odot\)). Significant He-burning may begin before the envelope is stripped. 
    \item Relax mass to 0.4 \(M_\odot\). Helium burning will start, or continue if started before mass stripping. Therefore, the model represents a subdwarf B (sdB) star formed following the CE (e.g., \citealt{xiong_subdwarf_2017}). The model runs through the sdB phase until it becomes a CO WD. 
\end{itemize}

Different He core masses will leave different masses of initial H on the model. Excess H will burn off during the sdB phase, meaning there is a maximum amount of H possible when the model enters the WD cooling track. For example, a 0.4 \(M_\odot\) He star starting with either a 0.39 \(M_\odot\) He core or a 0.395 \(M_\odot\) He core will wind up with roughly the same mass of H, despite starting with different values. This sets the maximum amount of H used in our grid, which is a few $\times$ $10^{-4}$ \(M_\odot\) in the WD mass range of interest. The initial mass of H can affect the duration of the sdB phase, but the effect is small compared to the overall duration.
Conversely, CO models with the lowest amount of H correspond to sdB models starting with low H.
Similarly to He-core WD models, the exact lower limit does not affect the WD cooling uncertainties and is not important for age estimates.

Our grid of CO WD models has its upper mass bound set by the masses of the primaries we model. Considering WDs 1$\sigma$ above measured values, that upper limit is 0.47 \(M_\odot\) from Table \ref{tab:binary parameters}. The lower limit is set by the transition between a He- and a CO-core WD, which depends on progenitor mass  (Sec.  \ref{progenitormass}), but reaches a low of $\approx \! 0.32$ \(M_\odot\)  (e.g., \citealt{moroni_very_2009}). 


\subsection{sdB + CO WD Behavior}

\label{sdB}

\begin{figure}
    \centering
    \includegraphics[scale=0.51]{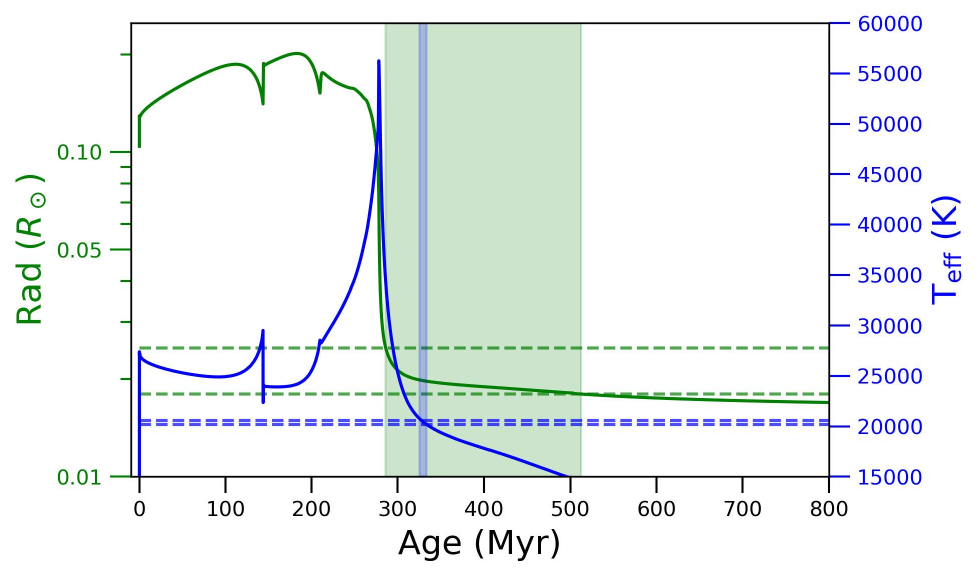}
    \caption{Similar to Fig. \ref{fig:cooling}, in blue (green) is the temperature (radius) curve of a 0.42 \(M_\odot\) sdB star which subsequently cools as a CO WD. Dashed lines show temperature/radius constraints of ZTF J1749's primary. In this case, temperature is the only useful constraint on the model's age. The sdB phase occurs during the first $\sim$300 Myr.}
    \label{fig:sdb_match}

\end{figure}

During the sdB phase defined by He-burning, the models stay at an elevated temperature of $T \! \sim \! 30,000 \, {\rm K}$ and radius of $R \! \sim \! 0.1 \, R_\odot$. During this phase, the star can exhibit "oscillations" in luminosity, temperature, radius, and nuclear burning power. This is due to He-burning shell flashes, as explained in  \cite{iben_relaxation_1986,sweigart_gravonuclear_2000,prada_moroni_very_2009}.
The sdB phase can last for a few times $10^{8}$ years, meaning that there is a significant time gap between the initial period of mass loss that forms the sdB (in our case, assumed to be CE evolution) and the formation of the CO WD. Once He burning has subsided, the star will cool and contract on to the WD cooling branch, with a CO-core. 


Our models' radii during the sdB phase are far too large to match to any of our WD primaries, and they only match during the WD phase. The temperature and radius cooling curves can then be compared to the constraints on the primary, and the cooling age turned into a birth period, in the exact same manner as performed for He WDs in Sec. \ref{binary cool}.  Fig. \ref{fig:sdb_match} shows a typical matching model for one of our binaries, including both the sdB phase and the WD cooling phase. Similarly to Sec. \ref{Nelemans WDs sec}, we instead apply temperature and surface gravity constraints for WD 0957-666 and WD 1101+364.

Table \ref{tab:CO result tab} presents a summary of the cooling ages/birth periods for each WD primary, assuming it is a CO-core WD formed from a sdB star. The one system not shown in the table (SDSS J0822) had no matching CO-models; unsurprisingly, it had the lowest-mass primary (mean value of 0.3 $M_\odot$), which is unlikely to be a CO WD.

\subsection{Comparison to He-core Models} \label{comp-He_CO}

Even for models with the same mass and same envelope mass, there are cases where there are good fits for He WD models, but not for CO WD models. Specifically, the radius and temperature cooling curves can be altered based on the model's composition.  Most noticeably, the radius of our CO models are always smaller than the radius of a corresponding He model. If the WD cores were purely degenerate, this would be unexpected - however, CO-core models are slightly smaller because of their heavier compositions and not being fully degenerate for the low masses we include. This is similar to the trend found in \cite{prada_moroni_very_2009}, where CO models had lower radii than He models when compared at the same temperature. In addition, \cite{prada_moroni_very_2009} found that CO WD models cooled faster (after the sdB phase) than He WD models.



\subsection{Effect of mass transfer during sdB phase}

Because the sdB star is at an elevated radius, it is possible that it undergoes mass transfer with its companion before ending He burning and contracting into a WD. We modeled conservative mass transfer as in Sec. \ref{flashes}, focusing on models with the most massive H envelopes during the sdB phase, as these are the most bloated. As expected, mass transfer strips much of the H envelope. However, the cooling behavior after the sdB phase (when mass transfer ends) is similar to models without mass transfer. This is because, even without mass transfer, most of the H envelope is burnt to He. Therefore, the H envelope will wind up being small ($\lesssim$ few $\times$ 1e-4 $M_\odot$) in either case. As long as mass transfer only strips the H envelope, it will likely not affect the parameter space of possible H mass on a cooling CO WD. If mass transfer resulted in mass loss from the He core, then it \textit{would} likely affect the cooling behavior \citep{bauer_phases_2021}. However, because we model DA WDs, we do not consider this possibility.

We also find that mass transfer of the H envelope does not significantly change the duration of the sdB phase (which is predominantly determined by the burning of He deeper in the star). Therefore, we find that mass transfer does not significantly change $T_{\rm{eff}}$ or $R$ as a function of age after the sdB phase. In principle, mass transfer affects the orbital period evolution and therefore the inferred birth period.
However, the effect is not very large  because the amount of transferred mass is relatively small.
Therefore, for our results, we run a grid of sdB models/CO WDs that do \textit{not} undergo mass transfer, as the grid's behavior will likely not be significantly altered by mass transfer.

\section{RGB progenitor models }

\label{progenitor main}

\subsection{Possible progenitor masses for CE evolution}

\label{progenitormass}

\begin{figure}
    \centering
    \includegraphics[scale=0.3]{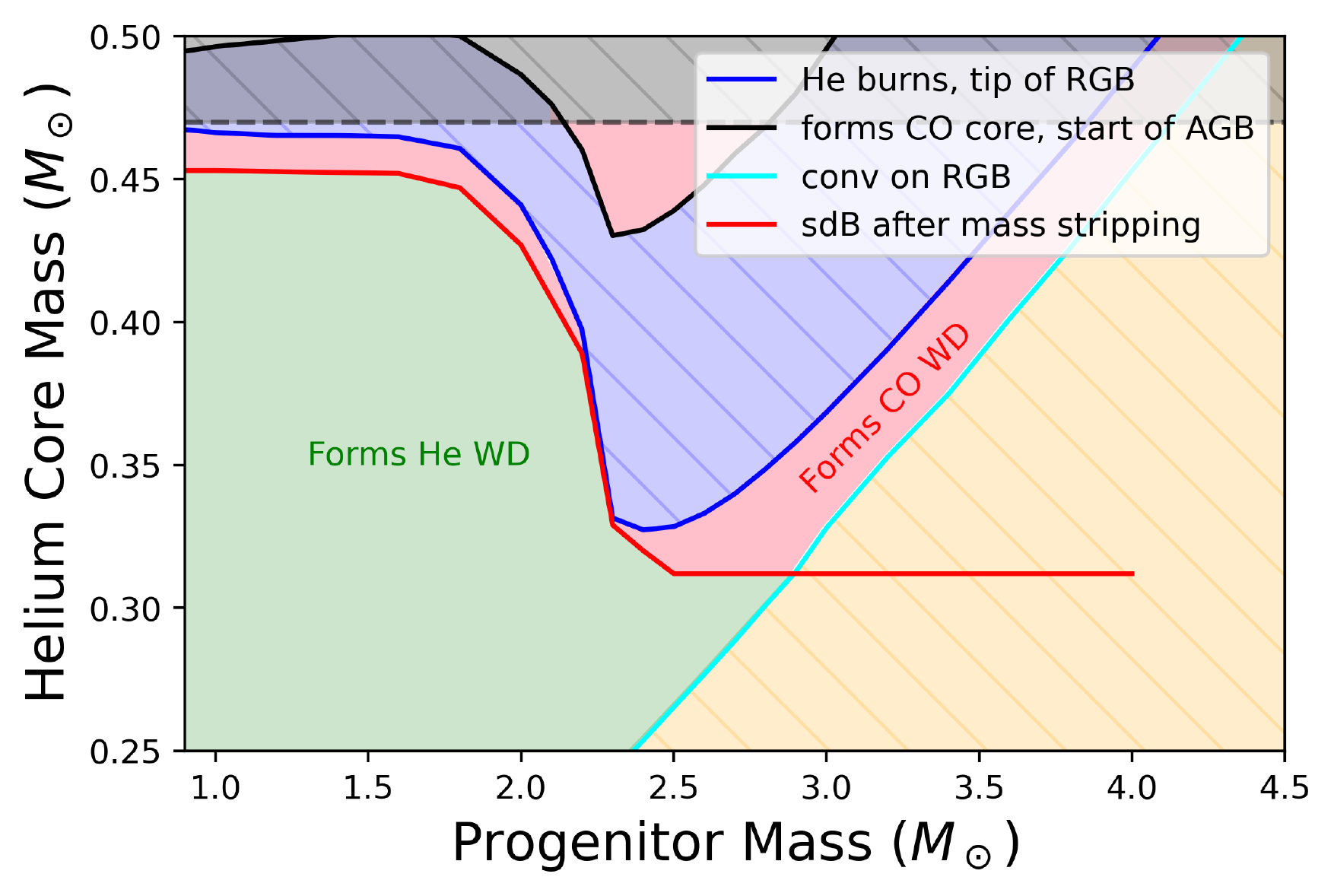}
    \caption{A chart showing the range of He/CO core masses that can be formed, as a function of progenitor mass. A given progenitor model moves upward on this diagram as it evolves.
    The horizontal gray region is above the mass of WDs relevant to our grid. The cyan curve represents the core mass where the model is half-convective by mass or radius - only progenitors above this line will undergo unstable mass transfer, and hence we do not consider WDs from progenitors in the yellow region. The red line is the core mass at which He burning will start following envelope stripping, while the blue line represents the He core mass at the tip of the RGB. The black line represents the formation of a CO core and the start of AGB ascent. In the shaded blue region, the star is smaller than its radius at the tip of the RGB, so a CE event  in this region is unlikely.
    The shaded green region is the possible domain of He WD progenitors, while 
    the shaded pink region is the possible domain of CO WD progenitors.
    }
    \label{fig:progenitors}
\end{figure}

For each matching WD model from the previous two sections, we determine the range of possible pre-CE progenitor masses and evolutionary states. Fig. \ref{fig:progenitors} summarizes the possible channels to creating WDs through CE evolution. The basic idea is that as a star evolves, its core mass increases and it moves vertically up the chart. It then encounters different regions that determine whether it can undergo CE evolution and the resulting composition of the WD. The core mass of the progenitor star is almost exactly equal to the mass of the WD that is formed after the envelope is stripped.
We do not consider progenitor masses below 0.9 $M_\odot$ as these stars will not have evolved off the main sequence within a Hubble time. 
This statement is slightly dependent on metallicity (e.g., \citealt{bazan_metallicity-dependent_1990}) but very low metallicities are unlikely for the field stars we consider. 

Our first criteria for successful CE evolution is that the progenitor star should be predominantly convective in order for unstable mass transfer to occur \citep{hjellming_thresholds_1987, soberman_stability_1997}. We find the mass of the He core at the point the star becomes half-convective by mass or radius (whichever comes first, but they occur almost at the same core mass), which is shown by the cyan line in Fig. \ref{fig:progenitors}. Above this line, the RGB star will be mostly convective and CE evolution allowed; below it, CE evolution is likely not possible. For example, a 2.5 $M_\odot$ star with a 0.25 $M_\odot$ core will not have a thick convective envelope; however, when its core increases to 0.3 $M_\odot$ it will have become mostly convective, and be eligible to undergo CE evolution. A more accurate criterion for CE could be obtained by analyzing whether mass transfer is stable using detailed models (e.g., \citealt{pavlovskii_stability_2017}). Their results show that mass transfer becomes much more unstable when the star develops a convective envelope after the main sequence. This partially justifies our choice, but a more thorough analysis would need to consider both the evolutionary state and the mass ratio of the system.


CE evolution will occur either on the RGB or AGB as the star is expanding. Between the RGB and AGB, the contracting star will not overflow its Roche-lobe if it has not done so already. We define the tip of the RGB by a spike in He-burning power (or equivalently a decrease in radius). The mass of the He core at the tip of the RGB is plotted in blue in Fig. \ref{fig:progenitors} - this curve is similar to one found in \cite{bauer_phases_2021}. We define the start of the AGB by the formation of a CO-core.  The He core at the start of the AGB is plotted in black. Therefore, the blue shaded region between blue and black lines is another forbidden region for CE evolution when the star is contracting.
For example, as a 2.5 $M_\odot$ star's He core evolves from 0.34 to 0.43 $M_\odot$ and it moves through the blue region, it is not likely to undergo CE evolution.

Finally, we consider where mass stripping representing CE evolution will create a He-core WD versus a CO-core WD. The transition between these scenarios is shown by the red line in Fig. \ref{fig:progenitors}. For a range of progenitor masses, we strip mass as in previous sections, creating a limited set of models. If the He-burning luminosity is negligible and the model cools similarly to those constructed in Sec. \ref{create_He}, it becomes a He WD. If He ignites before or after the mass stripping, leading to a sdB phase as in Sec. \ref{sdB}, it will become a CO WD. Above the red line, any model formed through mass stripping will form a CO WD; below it, a He WD. As a 2.5 $M_\odot$ progenitor moves upward across the red line, it will go from being able to form a He WD to a CO WD when its core mass exceeds $\approx 0.32 \, M_\odot$. In a couple of  cases, the He-burning power spikes sharply after mass-stripping and leads to numerical difficulties, but for most models we can track the formation of the CO-core as well.

In summary, the region of Fig. \ref{fig:progenitors} in which the progenitor will form a He WD through CE evolution is defined as 1) the star is mostly convective, 
2) helium has not ignited when the envelope is stripped, and it will not ignite afterwards. This region is shaded green in Fig. \ref{fig:progenitors}. There is an obvious transition around 2.2 $M_\odot$ where the parameter space to form He WDs drops sharply - this coincides with the boundary of the star's He core being degenerate or not \citep{nelemans_reconstructing_2000}.

The region of Fig. \ref{fig:progenitors} that can form a CO WD through CE evolution is defined as 1) the progenitor is mostly convective, 2) the progenitor is in an expansion phase below the tip of the RGB or above the start of the AGB, 3) helium has started burning (or will start burning) when the envelope is stripped. These regions are shaded in pink in Figure \ref{fig:progenitors}. There is a wide band above 2.3 $M_\odot$, where a CO WD can form with as low a mass as $\sim \! 0.32 \, M_\odot$. There is a smaller region between the red and blue lines from 0.9 to 2 $M_\odot$ - these represent models that undergo CE evolution very close to the tip of the RGB, resulting in the formation of canonical $\simeq 0.47 \, M_\odot$ sdB stars (e.g., \citealt{xiong_subdwarf_2017}). Finally, there is a transitional region from about 2 to 2.3 $M_\odot$.

\subsection{Determination of $E_{\rm bind}$}

\begin{figure}
    \centering
    \includegraphics[scale=0.55]{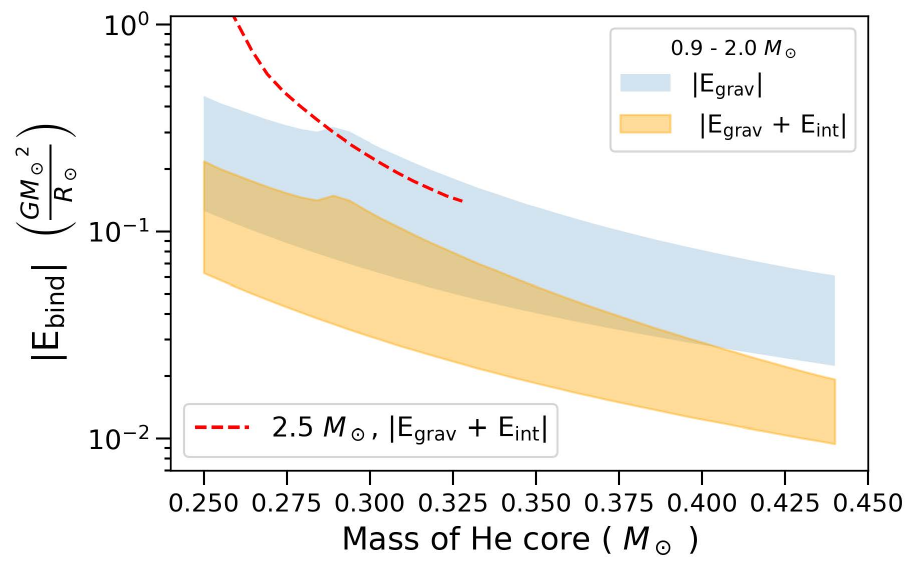}
    \caption{Envelope binding energy of RGB stars versus the helium core mass. Since $E_{\rm bind}$ is negative, a higher magnitude represents a more bound envelope. We plot the envelope's gravitational energy (blue shading), and gravitational + internal energies, with internal = thermal + recombination (orange shading). The range of $E_{\rm bind}$ corresponds to the possible range for 0.9 - 2.0 $M_\odot$ RGB stars, with the upper boundary from a 2.0 $M_\odot$ model and the lower from a 0.9 $M_\odot$ model. Also shown is $E_{\rm bind}$ for a 2.5 $M_\odot$ model (red dashed line), shown only up to a core mass of 0.32 $M_\odot$ (the tip of the RGB at that mass).}
    \label{fig:ebind_fig}
\end{figure}


\label{Ebind}

The CE event involves the unbinding of the donor star's envelope, which requires a source of energy to occur. The binding energy of the envelope in its simplest form is defined to be the gravitational binding energy $E_{\rm grav}$  alone, 
\begin{equation} \label{grav}
      E_{\rm grav}  =  \int_{M_{\rm{core}}}^{M_{\rm{tot}}} \frac{-G m(r) dm}{r} \, ,
\end{equation}
where $m(r)$ refers to the enclosed mass at some radius $r$. Unless noted, we define the mass of the He core $M_{\rm{core}}$ to be where $X_{\rm{H}}$ < 0.01 and $X_{\rm{He}}$ > 0.1 (mass fraction of H drops below 0.01 and mass fraction of He rises above 0.1). We investigate the effect of changing the core boundary definition in Appendix \ref{Core boundary effect} and find it to be small compared to other uncertainties. For all RGB models, we do \textit{not} include mass loss through winds. Including winds may decrease the progenitor mass by up to $\sim$10 per cent \citep{de_marco__2011} at the time of the CE event, but is a small effect relative to other uncertainties.

There are two other sources of energy that can make $E_{\rm bind}$ lower in magnitude, and the envelope easier to eject: the envelope's internal energy $E_{\rm int}$, and the energy released from recombination of elements within the envelope such as H and He \citep{han_possible_1994}. The internal energy of the envelope $E_{\rm int}$ is simply an integral of the internal energy per unit mass $U$, over the mass of the envelope:
\begin{equation} \label{int}
     E_{\rm int} =  \int_{M_{\rm{core}}}^{M_{\rm{tot}}} U dm \, .
\end{equation}
MESA defines internal energy $U$ as a sum of thermal energy and recombination energy \citep{Paxton2018} and both energy sources are therefore included in our definition of $E_{\rm int}$. The total binding energy is 
\begin{equation} \label{tot}
     E_{\rm bind} =  E_{\rm grav} + E_{\rm int} \, .
\end{equation}
With $E_{\rm grav}$  defined as negative, $E_{\rm int}$ is a positive quantity, and therefore makes the total $E_{\rm bind}$ less negative, i.e. the envelope less strongly bound. Here, we assume that \textit{all} the thermal energy contributes to unbinding the envelope, i.e. there is a thermal efficiency of one \citep{dewi_energy_2000}.

The behavior of $E_{\rm bind}$ is shown in Fig. \ref{fig:ebind_fig}, and depends both on the total mass of the RGB star and the mass of its core. We focus on RGB masses between 0.9 and 2.0  $M_\odot$ because they correspond to our main grid of He WD models. For a given progenitor mass, the envelope binding energy decreases as the core grows in mass and the RGB star expands. For a given core mass of a star ascending the RGB, the envelope radius is nearly independent of total mass. Hence, the value of $E_{\rm bind}$ is roughly proportional to the total mass squared, spanning a factor of $\sim$4 for $0.9-2\, M_\odot$ stars.




\subsection{Effect of progenitor mass on WD models}


\label{progenitor effect}

The progenitor mass used to actually create our grid of WD models can potentially affect that grid and its cooling behavior. However, we generally find that the effect is small. Therefore, the main uncertainty associated with the mass of the progenitor is through the envelope binding energy (previous section), and \textit{not} the WD cooling.

\subsubsection{He WDs}

As an example, a 0.3 $M_\odot$ He WD can be formed through a CE from a 0.9 $M_\odot$ to $\approx$ 2.8 $M_\odot$ progenitor RGB star (referring to Fig. \ref{fig:progenitors}). 
However, it would be time-consuming and redundant to create a grid of WD models from each possible progenitor mass. As mentioned in Sec. \ref{create_He}, the cooling behavior of He WDs with a  0.9 - 2.0 $M_\odot$ progenitor are similar  and we use a 1.2 $M_\odot$ progenitor model to create these WDs in MESA. However, models created from 2.1 - 2.8 $M_\odot$ progenitors show more significant cooling differences, likely because the core is less degenerate when formed (\citealt{nelemans_reconstructing_2000} uses 2.3 $M_\odot$ as the transition to non-degeneracy). See further discussion in Appendix \ref{He progen}.



\subsubsection{CO WDs}

\label{CO progenitor affects WD}

As an example, a 0.4 $M_\odot$ CO WD can be formed through a CE from two distinct regions in Fig. \ref{fig:progenitors}: either the pink region centered around a 3.5 $M_\odot$ progenitor, or a narrow region between the blue and red lines at $\approx$ 2.1 $M_\odot$. The latter region represents a sdB star that forms from mass loss very close to the tip of the RGB. 

Our main grid of sdB/CO WD models were created in MESA from a 3.0 $M_\odot$ progenitor. We compared models from the 3.0 $M_\odot$ progenitor, to models created from 1.0, 2.0. 3.5, and 4.0 $M_\odot$ progenitors. In general, the main effect is the duration of the sdB phase, not the behavior of the cooling CO WD. In most cases, the change in duration was small compared to the overall duration. The change was more significant when comparing canonical sdB models (creating $\sim$ 0.47 $M_\odot$ WDs)  from the 1.0  $M_\odot$ progenitor. However, such massive CO WDs were not matches to any of our systems (Table \ref{tab:CO result tab}), so they are likely irrelevant. See Appendix \ref{CO progen append} for more details.

\subsection{Matching WD models to progenitor star}

\label{match to progen}

For each matching WD model with mass $M_1$, we connect that model to the allowed regions of Fig. \ref{fig:progenitors} that could have formed it. We do so by assuming that the CE event occurs at the  point where the progenitor star's core mass equals the WD's initial core mass. 
Progenitors of 1.7 $M_\odot$ and below can create any of our He WD models (which go up to 0.45 $M_\odot$ in our grid). Higher mass progenitors can only form lower mass He WD models as shown in Fig. \ref{fig:progenitors}.

For a given progenitor mass and WD mass, we can therefore calculate $E_{\rm bind}$ at the exact point that the progenitor has a core mass equal to the mass of the WD. However, we do not calculate $E_{\rm bind}$ as the binding energy of the entire envelope, because our WD models also have a finite H envelope (i.e. not all of the progenitor's H envelope is unbound). We therefore integrate Eqs. \ref{grav} and \ref{int} from $M_{\rm{wd}}$ to $M_{\rm{tot}}$, where $M_{\rm{wd}}$ is slightly larger than the core mass. This reduces the associated $E_{\rm bind}$ for some WD models that have a relatively massive H envelope.


\section{Results}

\label{results main}

\subsection{Birth periods} \label{Birth periods all}

\begin{figure}
    \centering
    \includegraphics[scale=0.23]{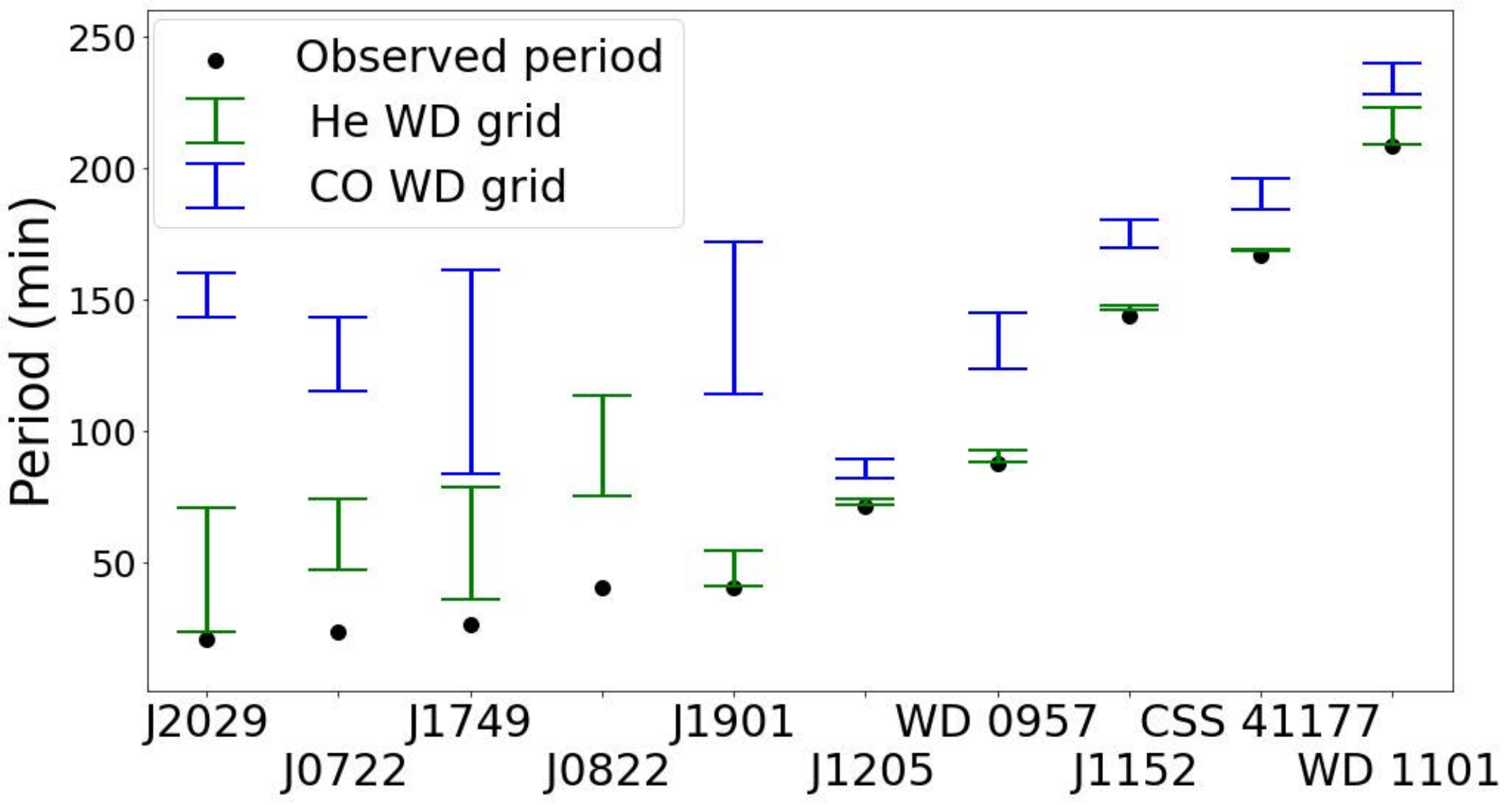}
    \caption{ The range of possible orbital periods at birth (immediately following the CE) for each of our systems, assuming the primary is a He WD (green points) or a CO WD (blue points). SDSS J0822 has no matching models consistent with a CO WD.}
    \label{fig:birth period}
\end{figure}

After finding the range of cooling times consistent with observations, we found the range of possible birth periods for each system that we model. Assuming inspiral due to GW emission only and integrating Eq. \ref{eq:inspiral} for possible $M_1$ and $M_2$ determines birth period for a given cooling age. The range of possible $M_1$ is from our matching WD models (i.e. Tables \ref{tab:matching_models} and \ref{tab:CO result tab}), which show good agreement with values in the literature. For most systems, the range of $M_2$ is solely from published literature measurements and uncertainties (i.e., the range in Table \ref{tab:binary parameters}). For WD 0957-666 and WD 1101+364, the mass ratio $q$ is instead used to find the range of $M_2$ values for each $M_1$. Fig. \ref{fig:birth period} shows the corresponding birth periods for both our main grids: He-core WDs created from a 1.2 \(M_\odot\) progenitor and CO-core WDs created from a 3.0 \(M_\odot\) progenitor.

Note there are no matching CO models for SDSS J0822 because the primary mass $M_1$ in this system is relatively low (mean value of 0.3 \(M_\odot\)) and most CO WDs form at higher masses. In contrast, all the other systems (with higher $M_1$) have matching CO models.  Because of the sdB phase which follows the CE (which can last 100s of Myr), our matching CO models are older than He models  and correspondingly give higher estimates for the system's birth period. The maximum birth periods we find correspond to orbital separations on the order of only 1 \(R_\odot\) - therefore, the overall assumption of CE evolution taking place is valid, since that separation is much less than the radius of an evolved RGB star.

The possible birth periods sometimes span a wide range due to uncertainties in cooling age, which results from models with a range of masses and hydrogen shell masses being consistent with the observed properties of a primary. In general, the uncertainty in hydrogen mass contributes much of the uncertainty in birth period.
For example, for ZTF J2029 the total possible birth period range (considering all primary masses) is 23 - 71 minutes, but 0.32 \(M_\odot\) models alone have a wide associated range (34 - 63 minutes). This latter range is \textit{not} predominantly from  uncertainty in $M_2$ (fixing $M_2$ only reduces the range by about 3 minutes). 
Instead, it is mostly due to varying cooling ages for models with different masses of hydrogen.

Several systems have a smaller range of possible birth periods.
SDSS J1152 and CSS 41177 have similar age uncertainties to other systems (see Table \ref{tab:matching_models}), but they were discovered at longer orbital periods. Hence, they cannot have been born at periods much greater than observed, because GW-driven orbital decay is less efficient at long periods. Similarly, the narrow birth period range for SDSS J1205 (the WD-BD binary) arises from the small mass of the BD companion, which decreases the GW orbital decay rate such that the observed period is very close to the birth period. For WD 1101, despite being observed at a long period, the cooling ages can be quite large (due to the system's low $T_{\rm{eff}}$), so there is a reasonable uncertainty in birth period. Conversely, the cooling ages are small for WD 0957 because of the WD primary's high $T_{\rm{eff}}$ of 30,000 K, as models of all masses quickly cool below 30,000 K, leading to a small birth period uncertainty.

\subsection{$\alpha_{\rm CE}$ Constraints}
\label{alpha_fin}

\subsubsection{He WD Simulations} 

\label{He main section}

   
 





The CE efficiency factor, $\alpha_{\rm CE}$, is defined via an energy parametrization as
\begin{equation} \label{eq:alpha}
     E_{\rm{bind}} = \alpha_{\rm{CE}} \left(  -\frac{G M_1 M_2}{2 a_{\rm{f}}} +\frac{G M_{\rm{i}} M_2}{2 a_{\rm{i}}}  \right) \, .
\end{equation}
Here, $M_{\rm{i}} $ is the mass of the progenitor star, and $a_{\rm{i}}$/$a_{\rm{f}}$ are the initial/final orbital separations. The value of $\alpha_{\rm CE}$ represents the fraction of the change in orbital energy that is used to unbind the donor star's envelope \citep{ivanova_common_2013}. An $\alpha_{\rm CE}$ of unity implies that all the orbital energy released by the inspiral goes into unbinding the hydrogen envelope (full conservation of energy). Values smaller than unity imply that some energy is lost during the CE event (e.g., because it is radiated away) or that the ejected material has positive kinetic energy when it escapes to infinity. Values greater than unity can only be achieved with an extra source of energy, e.g., energy released due to accretion on to the companion during the CE.

Equation \ref{eq:alpha} is a parameterization that relates the initial and final orbital energy to the binding energy of the progenitor star at the ``onset" of the CEE. Our method defines this onset to be when Roche lobe overflow first occurs. In principle, there could be a phase of stable mass transfer preceding or following the dynamical CEE (e.g., \citealt{ge_2010}). Our inferred value of $\alpha_{\rm CE}$ thus describes the orbital decay resulting from the entire process, but not its individual phases.

For each matching WD model (with mass $M_1$ and birth separation $a_{\rm{f}}$), we determine the corresponding value of $\alpha_{\rm CE}$ as follows. To link pre- and post-CE states, we map the newly formed WD model to a red giant progenitor model with that same core mass (Section \ref{match to progen}). Then, $E_{\rm bind}$ is computed using equations in Sec. \ref{Ebind}, where the energies are integrated over the ejected envelope. The companion mass $M_2$ is assumed to remain constant throughout the CE event. It is also assumed constant during any further phases of mass transfer (e.g., a H flash of the primary, or a bloated sdB phase).
    
The initial semi-major axis $a_{\rm{i}}$ is calculated by assuming the donor, which goes on to form the primary WD, is just overflowing its Roche lobe. We use the Roche lobe approximation of \citep{eggleton_aproximations_1983}
\begin{equation} \label{eq:roche}
   \frac{R_{\rm{i}}}{a_{\rm{i}}} = \frac{0.49q_{\rm{i}}^{2/3}}{0.6 q_{\rm{i}}^{2/3} + \ln{\left(1+q_{\rm{i}}^{1/3} \right)}} \, ,
\end{equation}
where the mass ratio is $q_{\rm{i}} \equiv M_{\rm{i}}/M_2$, and $R_{\rm{i}}$ is the radius of the progenitor star.
In practice, the initial orbital energy $G M_{\rm{i}} M_2/(2 a_i)$ does not matter greatly, since the final orbital energy is much greater in magnitude. For each possible configuration ($M_1$, $M_2$, etc.) of each system, we compute the corresponding value of $\alpha_{\rm CE}$ from equation \ref{eq:alpha}. We include internal energy (thermal plus recombination) into $E_{\rm bind}$ for our best estimate.


\begin{figure}
    \centering
    \includegraphics[scale=0.25]{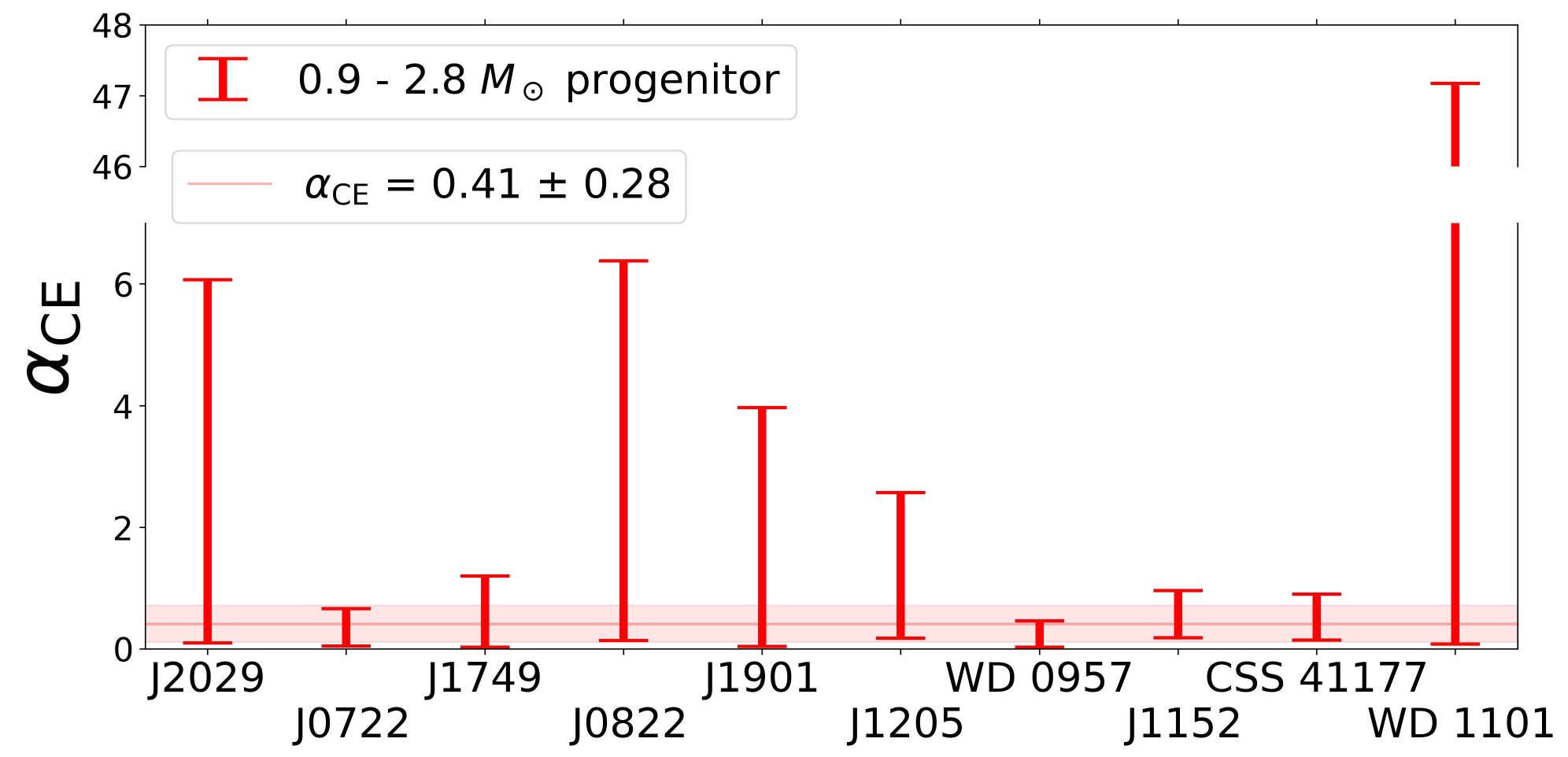}
    \caption{ Values of $\alpha_{\rm CE}$ using a grid of He WDs and progenitor mass from 0.9 - 2.8 \(M_\odot\) (the maximum mass RGB star that can produce a He WD from Fig. \ref{fig:progenitors}). The horizontal line and shaded region denotes a least squares fit to the ensemble of systems. }
    \label{fig:alpha_all}
\end{figure}

\begin{figure}
    \centering
    \includegraphics[scale=0.33]{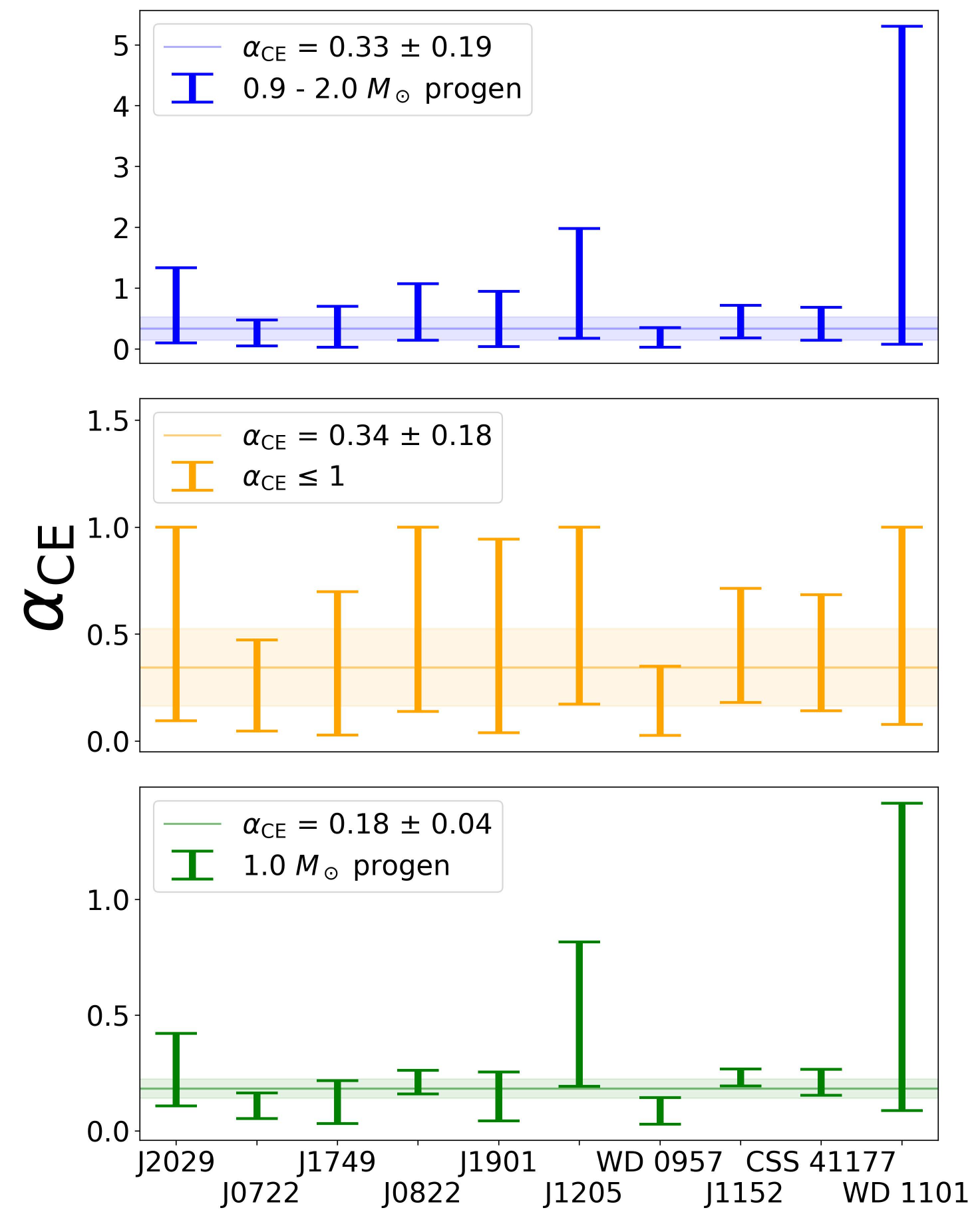}
    \caption{ Values of $\alpha_{\rm CE}$ for He WDs if progenitors of 0.9 - 2.0 \(M_\odot\) are assumed (top),  if $\alpha_{\rm CE}$ values above unity are discarded (middle), and if a progenitor of fixed mass 1.0 \(M_\odot\) is assumed (bottom). The shaded regions denote least squares fit to each ensemble of systems. }
    \label{fig:alpha_subplot}
\end{figure}

The possible range in $\alpha_{\rm CE}$ for each system, assuming the primary is a He WD, is shown in Fig. \ref{fig:alpha_all} and Fig. \ref{fig:alpha_subplot}, for different assumptions involving progenitor mass. Results for our grid of CO models  are discussed in Section \ref{alpha CO}. The results of Fig. \ref{fig:alpha_all} correspond to our most general criteria, in which viable He WD-progenitor pairings are made using Fig. \ref{fig:progenitors}, making the effective upper limit 2.8  \(M_\odot\). In addition, progenitors more massive than 2.2 $M_\odot$ cannot form a He WD more massive than about 0.32 $M_\odot$. Therefore, half of the systems we model, with relatively high-mass He WDs, cannot have progenitors more massive than 2.2 $M_\odot$. Note that we use our main grid of WD models, constructed in MESA from a 1.2 $M_\odot$ progenitors, for our calculations (see Appendix \ref{He progen} for further discussion).



All systems are consistent with CE efficiencies $\alpha_{\rm CE} < 1$ when considering typical progenitors of $\sim$1-2 $M_\odot$. For several systems, $\alpha_{\rm CE}$ values in Fig. \ref{fig:alpha_all} can be substantially above unity when considering massive donors. For example, ZTF J2029 and WD 1101 have extremely high associated $\alpha_{\rm CE}$ values because their low-mass He primaries can be formed from RGB stars up to 2.8 \(M_\odot\), leading to high values of $E_{\rm bind}$. In contrast, systems like WD 0957 and CSS 41177 have higher mass He primaries ($\sim \! 0.4 \ M_\odot$), limiting their progenitors to a maximum of 2.1 \(M_\odot\), and hence lower $E_{\rm bind}$ and $\alpha_{\rm CE}$.
If we assume a constant value of $\alpha_{\rm CE}$ applies to all systems, we calculate a value of $\alpha_{\rm CE} = 0.41 \pm 0.28$ using a least squares fit with each system weighted by the inverse of the possible range in $\alpha_{\rm CE}$ (i.e., the maximum - minimum value).

The most likely values of $\alpha_{\rm CE}$ are difficult to determine because they depend on the uncertain progenitor mass distribution, which may have been affected by a previous phase of stable mass transfer. High-mass progenitors are less likely because they require unphysically large values of $\alpha_{\rm CE}$ and are less likely according to the initial mass function. To estimate more realistic ranges $\alpha_{\rm CE}$, in the top panel of Fig. \ref{fig:alpha_subplot}, we restrict the progenitor mass to 0.9 - 2.0 \(M_\odot\).
In this case, possible values of $\alpha_{\rm CE}$ range from $\sim 0.1$ to $\sim 2$ for all systems except WD 1101, where $\alpha_{\rm CE}$ reaches a maximum over 5. If we again assume a constant value of $\alpha_{\rm CE}$ applies to all systems, we calculate a value of $\alpha_{\rm CE} = 0.33 \pm 0.19$ using the same method as above.

In the middle panel of Fig. \ref{fig:alpha_subplot}, we eliminate all values of $\alpha_{\rm CE}$ > 1 as they are not physical under this formalism without an extra source of energy. The resulting best-fit for $\alpha_{\rm CE}$ then becomes $\alpha_{\rm CE} = 0.34 \pm 0.18$. In the bottom panel of Fig. \ref{fig:alpha_subplot}, we instead assume a fixed progenitor mass of $1.0 \, M_\odot$ to demonstrate the small ranges in $\alpha_{\rm CE}$ that result. Assuming a $1.0 \, M_\odot$ progenitor results in a best-fit $\alpha_{\rm CE} = 0.18 \pm 0.04$, which represents an approximate lower limit to the actual CE efficiency.

In summary, 
we find that given reasonable assumptions on the progenitor, a universal $\alpha_{\rm CE}$ substantially below unity is favored. This conclusion is strengthened by the fact that several systems can only have low values of $\alpha_{\rm CE}$ due to their high-mass WDs, which require low-mass progenitors with low $E_{\rm bind}$
These several systems therefore have associated $\alpha_{\rm CE}$ values less than unity regardless of our assumptions regarding the progenitor.




\subsubsection{Uncertainties in $\alpha_{\rm CE}$}

For most systems, the uncertainty in $\alpha_{\rm CE}$ is a factor of several. Most of the range comes from uncertainty in the progenitor mass, which affects $E_{\rm{bind}}$ by a factor of $\sim$ 2 - 4 
when assuming progenitors of 0.9 - 2.0 \(M_\odot\) and by a greater factor if considering progenitors up to 2.8 \(M_\odot\) (Figure \ref{fig:ebind_fig}). In Fig. \ref{fig:alpha_subplot}, WD 1101 has the highest associated $\alpha_{\rm CE}$ values because of the combination of a low primary mass (minimum 0.27 \(M_\odot\), which results in high envelope binding energies), and a long birth period. The system with the second highest range of $\alpha_{\rm CE}$ values (SDSS J1205) has a BD companion, so the low values for $M_2$ drive $\alpha_{\rm CE}$ upward.

The second largest uncertainty comes from observational/modeling uncertainties in the primary mass $M_1$, even though it is usually a secondary uncertainty in regard to birth period. Because lower (higher) values of $M_1$ require more (less) compact RGB progenitors with higher (lower) binding energies, uncertainty in $M_1$ translates to a substantial uncertainty in $E_{\rm bind}$ and hence to $\alpha_{\rm CE}$. Additionally, referencing Table \ref{tab:matching_models} and Figure \ref{fig:birth period}, there is another factor of $\lesssim$ 2 uncertainty from max/min birth periods for these systems (although several have $\lesssim$ 5 per cent uncertainty) that arises primarily from the uncertainty in the WD's H shell mass.

The configurations that result in upper/lower values of $\alpha_{\rm CE}$ can be understood by examining equation \ref{eq:alpha}. Since the initial orbital energy term is usually small, we can ignore that term and solve for $\alpha_{\rm CE}$:
\begin{equation}
    \alpha_{\rm CE} \sim \frac{2 E_{\rm{bind}}}{G M_1 M_2}  \left({\frac{G (M_1 + M_2) {P_{\rm{f}}^2}}{4 \pi^2}}\right)^{1/3} \, ,
\end{equation}
where $P_{\rm{f}}$ is the final (birth) period.
Maximum values of $\alpha_{\rm CE}$ arise from a high progenitor mass and a low $M_1$ (creating high $E_{\rm bind}$). 
Although $P_{\rm{f}}$ tends to \textit{decrease} for lower $M_1$, that effect is not strong enough to dominate.
Minimum values of $\alpha_{\rm CE}$ are associated with models with a low progenitor mass and a relatively high $M_1$ (creating low $E_{\rm bind}$) . 

For illustrative purposes, we assume progenitors of 0.9 - 2.0 \(M_\odot\) in the following two paragraphs. In the case of ZTF J2029, the maximum value of $\alpha_{\rm CE}$  is associated with a 0.26 \(M_\odot\) WD model (the lowest mass consistent with observations), a 2.0 \(M_\odot\) progenitor(the highest mass considered in this example), and a birth period of 24 min. The minimum value of $\alpha_{\rm CE}$  is associated with a 0.35 \(M_\odot\) WD model, a 0.9 \(M_\odot\) progenitor, and a birth period of 55 min. Therefore, the maximum value of $\alpha_{\rm CE}$  counter-intuitively is associated with a model with one of the lowest birth periods, and vice versa.

Similarly, low uncertainties in birth period do not necessarily lead to tight ranges in $\alpha_{\rm CE}$.  ZTF J0722 has  a $\sim$ 50 per cent difference between lower/upper values of birth period, whereas  CSS 41177 has less than 1 per cent difference. However, the range in possible $E_{\rm{bind}}$ varies greatly for both (a factor of $\sim$5 for J0722 and a factor of $\sim$3.5 for CSS 41177), leading to high ratios of max/min $\alpha_{\rm CE}$ ($\sim$6 for J0722 and $\sim$4 for CSS 41177).
Hence, the benefit of  having a well-constrained birth period is largely wiped out by the uncertainty in $E_{\rm{bind}}$ due to the unknown progenitor mass. 

\begin{figure}
    \centering
    \includegraphics[scale=0.25]{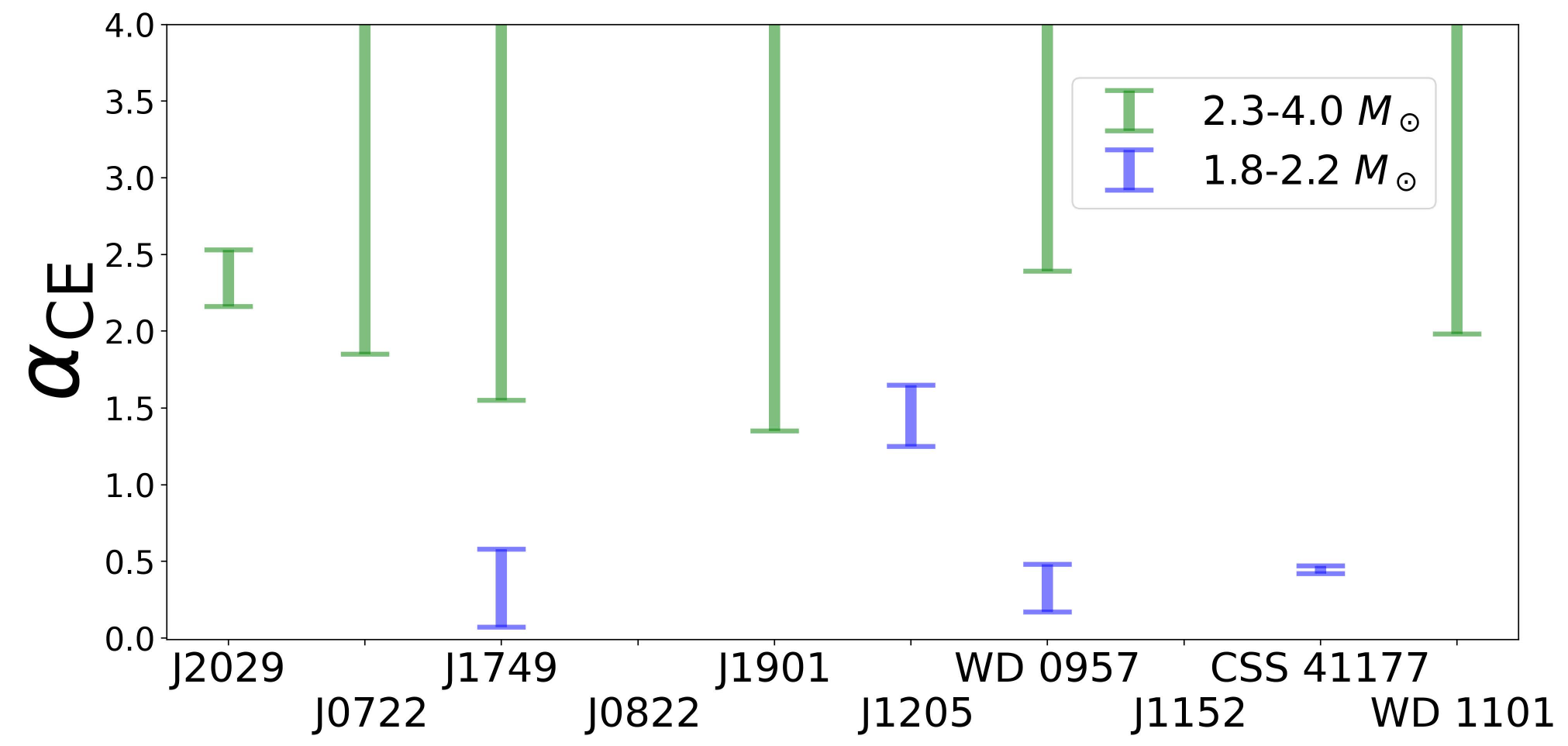}
    \caption{ $\alpha_{\rm CE}$ ranges assuming CO WDs arising from progenitors of the specified mass ranges. Not shown are green errorbars for J1205 (lower value $\sim 10$) and J1152 and CSS 41177 (lower values $\sim 4$). Otherwise, a lack of value indicates such a progenitor cannot create a relevant CO WD.}
    \label{fig:alpha_He_and_CO}
\end{figure}

\subsubsection{CO WD results}

\label{alpha CO}

From Fig. \ref{fig:progenitors}, the progenitors that can form a CO WD through CE evolution can be subdivided into 3 regions, 2 of which we focus on. For 1.8 - 2.2 $M_\odot$ progenitors, CO WDs of mass 0.39 - 0.45 $M_\odot$ can be created. Lower mass progenitors would create CO WDs more massive than 0.45 $M_\odot$, but those are not relevant to our matching CO models (Table \ref{tab:CO result tab}).
A 2.3 $M_\odot$ progenitor can only form CO WDs near 0.32 $M_\odot$, and for progenitors more massive than 2.3 $M_\odot$, the mass of viable CO WDs slowly rises. We therefore report $\alpha_{\rm CE}$ values separately for 
1.8 - 2.2 $M_\odot$ progenitors and 2.3 - 4.0 $M_\odot$ progenitors. We use the same grid of CO WDs in both cases, although such models were created with a 3.0 $M_\odot$ progenitor (justified in Sec. \ref{CO progenitor affects WD}). Similarly to previous sections, we map the CO WD models to progenitors in the regions of Fig. \ref{fig:progenitors} that could have formed them through CE evolution (in this case, the two lower pink regions).

The results for $\alpha_{\rm CE}$ are shown in Fig. \ref{fig:alpha_He_and_CO}. 
The $\alpha_{\rm CE}$ values for the progenitors with $M \! > \! 2.3 \, M_\odot$ are all greater than 1 and likely represent an unphysical pathway. The lowest values (still greater than 1) come from a 2.3 $M_\odot$ progenitor and CO models near 0.32 $M_\odot$. 
SDSS J0822 has no matching CO models. SDSS J1152, CSS 41177 and SDSS J1205 have models associated with progenitors of $M \! > \! 2.3 \, M_\odot$ but the corresponding values of $\alpha_{\rm CE}$ are exceptionally high, with minimum $\alpha_{\rm CE}$ values greater than $\sim$4.
The values are so high because these systems have high post-CEE birth periods and associated CO WD masses  $\gtrsim$ 0.35 $M_\odot$. From Fig. \ref{fig:progenitors}, the $M \! > \! 2.3 \, M_\odot$ progenitors that can form such a CO WD through CE evolution are at masses greater than 3 $M_\odot$, resulting in very high envelope binding energies and high  $\alpha_{\rm CE}$.

The results for 1.8 - 2.2 $M_\odot$ progenitors are more interesting. In the case of 5 systems, no results are shown because this set of progenitors is incapable of forming a CO WD with mass $\lesssim$ 0.39 $M_\odot$ through CE evolution, and these 5 systems only have matching CO WD models with masses below 0.39 $M_\odot$. Again, no result is shown for SDSS J0822 because it has no matching CO WDs of any mass.
ZTF J1749, SDSS J1205, WD 0957, and CSS 41177, in contrast, have matching CO WD models with masses $\gtrsim$ 0.39 $M_\odot$.
The first three of these can have $\alpha_{\rm CE}$ values less than 1, so a CO WD is possible. For SDSS J1205, the low mass of the BD companion drives $\alpha_{\rm CE}$ higher than 1, so a CO WD is unlikely. Assuming a constant value of $\alpha_{\rm CE}$ applies to all 4 systems and doing a fit in the same manner as in Section \ref{results main}, we find $\alpha_{\rm CE}$ = 0.46 $\pm$ 0.05, which lies in the range of our estimate from He WD models. 

The CO WD models with 1.8 - 2.2 $M_\odot$ progenitors have $\alpha_{\rm CE}$ values that overlap with the values from our main He WD grid. 
This may seem surprising, as CO models cool more slowly and entail significantly higher birth periods than He models. However, the behavior of $E_{\rm{bind}}$ can offset this effect. In the case of ZTF J1749 for example,  comparing to the top panel of Fig. \ref{fig:alpha_subplot}, the maximum value of $\alpha_{\rm CE}$ assuming a He WDis associated with a 0.32 $M_\odot$ WD, 2.0 $M_\odot$ progenitor. The minimum value of $\alpha_{\rm CE}$ assuming a CO WD is associated with a 0.45  $M_\odot$ WD model, 1.9 $M_\odot$ progenitor. Even though the birth period is roughly double for the CO model versus the He model, $E_{\rm{bind}}$ is reduced by a factor of $\sim 10$ on account of the \textit{higher} mass WD, which means the progenitor is higher up the RGB.

The large difference between results for 1.8 - 2.2 $M_\odot$  and $M \! > \! 2.3 \, M_\odot$ progenitors is similarly based on $E_{\rm{bind}}$.
Whereas lower mass progenitors ignite He burning at the tip of the red giant branch, higher mass progenitors ignite He burning closer to the base of the red giant branch when the progenitor is much more compact. This leads to a factor of $\sim$3 change in $E_{\rm{bind}}$, which results in a gap between the highest values of $\alpha_{\rm CE}$ for the 2.2 $M_\odot$ progenitor and the lowest values of $\alpha_{\rm CE}$ for the 2.3 $M_\odot$ progenitor.

\section{Discussion}

\label{discussion}

\subsection{Comparison to previous studies}

\cite{nelemans_reconstructing_2000} and \cite{nelemans_reconstructing_2005} used similar techniques to constrain CE efficiency for several WD binaries. They argued that observed short-period WD binaries require large orbital decay during the second phase of mass transfer, but not the first phase of mass transfer. In other words, they found that an energy formalism based on $\alpha_{\rm CE}$ applied to the first phase of mass transfer could not reproduce observed systems, so they argued for an angular momentum-based CE prescription (the $\gamma$-algorithm). We interpret their results as evidence that the first stage of mass transfer is usually stable (though perhaps not conservative), while the second stage of mass transfer is unstable. We have not modeled the first stage of mass transfer, so we defer a more thorough comparison to future work. Based on the modeling presented here, a nearly constant value of $\alpha_{\rm CE}$ in the second phase of mass transfer can reproduce the observed systems.

We can compare our results for WD 0957 and WD 1101 to the work by \cite{nelemans_reconstructing_2005}, where they present their results in terms of $\alpha_{\rm CE} \lambda$, defined as
\begin{equation} \label{lambda eq}
    \alpha_{\rm{CE}} \lambda \left(  -\frac{G M_1 M_2}{2 a_{\rm{f}}} +\frac{G M_{\rm{i}} M_2}{2 a_{\rm{i}}}  \right) =    \frac{G M_{\rm{i}} M_{\rm{env}}} {R} \, ,
\end{equation}
where $M_{\rm{env}}$ is the mass of the star's envelope.
In our models, for WD 0957 we find that $\alpha_{\rm CE} \lambda$ ranges from 0.02 to 0.50, whereas for WD 1101 $\alpha_{\rm CE} \lambda$ ranges from to 0.05 to 8.0. We interpret Fig. 5 of \cite{nelemans_reconstructing_2005} and estimate that the most likely range of  $\alpha_{\rm CE} \lambda$ for WD 0957, where many models overlap, is from about 0 to 0.5. This agrees well with our estimate.
For WD 1101, their most likely range of  $\alpha_{\rm CE} \lambda$ seems to be from about 0.2 to 0.8. If we restrict our WD mass range to 0.27 - 0.31 $M_\odot$, our $\alpha_{\rm CE} \lambda$ values reach a minimum of 0.3, roughly consistent with theirs. However, our possible range of $\alpha_{\rm CE}$ is much larger than \cite{nelemans_reconstructing_2005} because we consider a wider range of primary masses, corresponding to a wider range of progenitor envelope binding energies and therefore a wider range of $\alpha_{\rm CE}$.

\cite{zorotovic_close_2022} performed a similar analysis for several WD-BD binaries, finding that they are consistent with a low CE efficiency ($\alpha_{\rm CE}$  of about 0.2 - 0.4). They exclude recombination energy (but do include thermal energy), meaning their $\alpha_{\rm CE}$ values should be decreased slightly to compare with our values. For SDSS J1205, they found $\alpha_{\rm CE}$ to be between 0.18 and unity using progenitor masses of about 1.1 - 1.9 $M_\odot$.  For a similar range of progenitor masses, our value of $\alpha_{\rm CE}$ ranged between 0.19 and 2.
However, if we only use masses within 1$\sigma$ of the quoted values (as \citealt{zorotovic_close_2022} appear to have done), our $\alpha_{\rm CE}$ values reach a minimum of 0.4 instead of 0.19. Therefore our required $\alpha_{\rm CE}$ values appear to be somewhat larger than theirs, 
but we agree that $\alpha_{\rm CE}$ substantially less than unity can explain WD-BD binaries.
SDSS J1205 was also examined by \cite{parsons_two_2017}, who found $\alpha_{\rm CE} \lambda$ for this system ranges from about 0.1 to 0.6 for a progenitor mass ranging from 0.9 to 1.5 $M_\odot$ and a WD mass of 0.43 $M_\odot$ - we find $\alpha_{\rm CE} \lambda$ from about 0.1 to 0.67 for these same parameters. 
Our values are therefore in good agreement with theirs.

Our primary result for He WDs with 0.9 - 2.0 $M_\odot$ progenitors is that all systems are consistent with a CE efficiency of $\sim$ 0.2 to 0.4. In a study of post-CE binaries composed of a WD and a main sequence star, \cite{zorotovic_post-common-envelope_2010} determined ranges of $\alpha_{\rm CE}$ for 60 binaries. Note that they refer to recombination energy as internal energy, differing from our terminology.  When not including the recombination energy, they find $\alpha_{\rm CE}$ close to 0.5 works best to describe most systems \citep{zorotovic_2022}. Including some recombination energy would shift $\alpha_{\rm CE}$ to slightly lower values, particularly for progenitors on the AGB. Hence, their results are approximately consistent with our best-fit $\alpha_{\rm CE}$.

\cite{hernandez_white_2021, hernandez_white_2022} modeled post-CE binaries and found that $\alpha_{\rm CE}$ between 0.2 to 0.3 can reproduce binaries with a WD and a AFGK-type companion, without the inclusion of recombination energy.  
However, \cite{hernandez_white_2021, hernandez_white_2022} also found wider ranges of $\alpha_{\rm CE}$ ($\sim 0.1 - 1$) could be consistent with their systems. Because recombination energy is relatively unimportant for our RGB progenitors, their values of $\alpha_{\rm CE}$ can be compared directly to ours and appear consistent.

\cite{davis_is_2012}  reports $\alpha_{\rm CE}$ values for WD-main sequence binaries.
In their models including internal energy, $\alpha_{\rm CE}$ has a large scatter from about 0.02 - 2, but most systems are consistent with our inferred range of 0.2 - 0.4.
Similar to \cite{zorotovic_post-common-envelope_2010}, they find higher values of $\alpha_{\rm CE}$  for RGB progenitors versus AGB progenitors.
\cite{davis_is_2012} explain that their $\alpha_{\rm CE}$ values above 1 come from WD-BD binaries. For the single WD-BD binary that we modelled (SDSS J1205), we also find a median value of $\alpha_{\rm CE}$ slightly larger than 1, but the range for that system extends down to $\alpha_{\rm CE}$ $\sim$ 0.17. Hence, our results indicate brown dwarfs can only survive the CE when the WD's progenitor is relatively low-mass and high up the RGB. \cite{davis_is_2012} find that their $\alpha_{\rm CE}$ values \textit{decrease} with increasing progenitor mass and increasing $M_2$. 
Because of our smaller sample size, we did not perform a correlation between $\alpha_{\rm CE}$ and the parameters of each system.
 
For some of the same systems, \cite{de_marco__2011} found lower values of $\alpha_{\rm CE}$ than \cite{davis_is_2012}. \cite{de_marco__2011} included thermal energy in their estimates of the energy budget. Our inferred range of $\alpha_{\rm CE}$ ($\sim$ 0.2 - 0.4) is consistent with most of their results. They find a trend of \textit{increasing} values of $\alpha_{\rm De}$ for increasing progenitor masses.  However, they use initial to final mass relations to relate the WD mass to a progenitor mass, which may be problematic. For instance, a $0.55 \, M_\odot$ WD could be produced by a $1 \, M_\odot$ progenitor high on the AGB, or a $3 \, M_\odot$ progenitor soon after core He burning. It is possible their assumption introduces artificial trends into the results. When they modelled binaries that underwent CE evolution on the RGB, they used a progenitor mass of 1.19 $\pm$ 0.40 $M_\odot$, based solely on the initial mass function of \cite{kroupa_variation_2001}.



Future work should re-investigate whether there really is a trend between $\alpha_{\rm CE}$ and progenitor or companion mass. In our analysis, we find that there may be a trend where $\alpha_{\rm CE}$ is inversely correlated with the mass of the WD primary, $M_1$. However, this could be a result of correlations between uncertainties in $M_1$ and the inferred $\alpha_{\rm CE}$, because higher values of $M_1$ correspond to progenitors with smaller binding energies and hence lower inferred values of $\alpha_{\rm CE}$. More rigorous analysis including the correlations between the measurement and model uncertainties for each system should be the subject of future work. 


By analysing WD binaries in the ELM Survey, \cite{brown_most_2016} found that in order to match the observed number of short-period systems, most of the progenitors of He-CO WD binaries detach from the CE at an orbital period less than 1 hour. Four of the systems (ZTF J2029, J0722, J1749 and J1901) we have modelled were likely born at a period less than an hour, so our results are qualitatively similar to those of \cite{brown_most_2016}. However, the typical WD mass in the ELM Survey of $\sim 0.2 \, M_\odot$ is lower than the average mass of systems we analyse. Lower-mass WDs (which have progenitors with higher binding energies) are expected to be born at shorter periods on average, assuming the same $\alpha_{\rm CE}$, helping to explain the short birth periods they infer. 
However, we caution that a substantial fraction of low-mass He WDs ($M \! \lesssim 0.25 \, M_\odot$) are likely formed via stable mass transfer rather than through CE evolution \citep{li_formation_2019}, complicating their analysis.

\cite{han_2002} and \cite{han_2003} examined the formation of short-period sdB binaries through CE evolution, and found $\alpha_{\rm CE}$ of 0.75 to be the most appropriate to replicate the observed period distribution, but with only 75\% of thermal energy contributing to envelope binding energy. Therefore, their efficiency factor $\alpha_{\rm H}$ is defined as 

\begin{equation}
    \alpha_{\rm H} \Delta E_{\rm{orb}} = E_{\rm{grav}} + \alpha_{\rm H} E_{\rm{int}} \, .
\end{equation}
One can show that their efficiency factor can be related to ours by
\begin{equation}
    \alpha_{\rm H} = \frac{\alpha_{\rm CE}}{1 - \beta + \beta \alpha_{\rm CE}} \, .
\end{equation}
where $\beta=-E_{\rm int}/E_{\rm grav} \approx 0.5$ for stars low on the red giant branch. Hence $\alpha_{\rm H} \sim 0.75$ corresponds to $\alpha_{\rm CE} \! \sim \! 0.6$ by our definition, somewhat larger than our preferred value, though they note that $\alpha_{\rm CE}$ could not be accurately constrained. 

Recently, \cite{ge_2022} analyzed the CE ejection efficiency short-period binaries containing an sdB star with an M-dwarf or WD companion. A large fraction of their systems are consistent with $\alpha_{\rm CE} \sim 0.2-0.4$, in line with our results. However, some of their systems fall above or below these limits, and it is not clear why.

\cite{sandquist_2000} performed CE simulations for a few different progenitor masses, evolutionary states, and companion masses. Although their simulations only probed the dynamical phase of the CE and could not be run to complete envelope ejection, their estimates of $\alpha_{\rm CE} \sim \! 0.1-0.5$ during the initial phase of the CE are consistent with our empirical estimates. \cite{nandez_recombination_2015} used 3D simulations to construct DWD binaries, with low-mass red giant star progenitors from $\sim$1-1.8 $M_\odot$. They incorporated both thermal and recombination energy, and found an $\alpha_{\rm unb}$ of 0.2 - 0.44 (representing energy taken away by unbound material). This is related to the CE efficiency via $ \alpha_{\rm CE} = 1- \alpha_{\rm unb}$, leading to an $\alpha_{\rm CE}$ of 0.56 - 0.8. The minimum value of this range is slightly above the upper bounds of our fiducial range. Additionally, our maximum values correspond to high-mass progenitors, while their minimum values correspond to low-mass progenitors, so their values of $\alpha_{\rm CE}$ appear to be inconsistent with ours. This tension should be examined in future work.

Simulations by \cite{ohlmann_hydrodynamic_2015} with a 2 $M_\odot$ donor star only ejected 8 per cent of the envelope mass on a dynamical time-scale. They suggest that either processes on a thermal time-scale or the contribution of recombination energy may lead to full unbinding. Similarly, simulations by \cite{ricker_amr_2012} with a 1.05 $M_\odot$ donor star only ejected 26 per cent of the envelope mass, while the rest of the envelope remained bound.
Recently, \cite{law-smith_successful_2022} performed simulations with a 12 $M_\odot$ donor star where the envelope was completely ejected. They found $\alpha_{\rm CE} \approx$ 0.1 - 2.7 depending on how much material is ejected after the end of their simulations. 
However, they only included gravitational potential energy in their definition of envelope binding energy, meaning their values should be roughly halved to compare to ours, and therefore appear to be similar to ours.


\subsection{Constraining the Progenitor mass}

The main uncertainty in $\alpha_{\rm CE}$ for each system arises from the uncertain progenitor mass. If this could be constrained some other way, a more precise $\alpha_{\rm CE}$ could be determined. \cite{zorotovic_close_2022} uses the brown dwarf companions to estimate the total age of the system and therefore constrain the WD progenitor mass. Discovering a double WD binary in a cluster with known turnoff mass could also constrain the initial progenitor masses.
Similarly, discovering a WD binary in a widely separated tertiary \citep{toonen_evolution_2016} whose third star could be age-dated could constrain progenitor mass. 
A complication for both of these scenarios is that the primary (second-formed) WD progenitor may have accreted mass from its companion during a prior phase of stable mass transfer, increasing the mass of the primary WD progenitor and decreasing its lifetime. Indeed, the population synthesis of \cite{ruiter_2010} predicts that the vast majority of binary He WDs (which represent most of our sample) are formed in this manner.

\subsection{Other progenitor uncertainties}

\label{progenitor discuss}

The effect of metallicity on cooling He WD models is discussed in Appendix \ref{metallicityHe}. In the case of SDSS J0822, the discovery paper \citep{brown_discovery_2017} argues that the WD binary is likely located in the Galaxy's halo based on its distance. Therefore, it is possible that it has a significantly lower metallicity than solar. Assuming a tenth of solar metallicity for WD models increases the maximum H envelope mass, thus enlarging the maximum possible cooling age and increasing the maximum birth period for SDSS J0822 by about 35 per cent (while not significantly changing the minimum period).

Metallicity is also important in the progenitor RGB models, where changing metallicity changes the star's radius (and therefore $E_{\rm{bind}}$) as a function of its He core mass. The difference in $E_{\rm{bind}}$ (compared to the Z = 0.02 model) ranges from 20 per cent to 40 per cent for Z = 0.0067, and from 40 per cent to 80 per cent for Z = 0.002. These differences would propagate to $\alpha_{\rm CE}$. However, these uncertainties are still lower than those associated with the mass of the progenitor (where $E_{\rm{bind}}$ can triple). In the case of SDSS J0822, reducing metallicity for the progenitor star by a factor of ten increases $E_{\rm{bind}}$ by about 40 per cent. Together, the changes in birth period and $E_{\rm{bind}}$ lead to values of $\alpha_{\rm CE}$ approximately doubling. Using doubled $\alpha_{\rm CE}$  values for this one system does not substantially affect the least squares fit in Sec. \ref{results main},
but using lower metallicity progenitors to model all systems would substantially increase our best estimate of $\alpha_{\rm CE}$. However, it seems unlikely that a large fraction of these systems formed from low-metallicity progenitors (but see \citealt{thiele_applying_2021}).


For a given progenitor mass, parameters such as convective mixing length and convective overshoot can affect the relationships between main sequence mass, core mass, and red giant radius.
Decreasing the convective mixing length $\alpha_{\rm{MLT}}$ has been found to decrease the temperature and increase the radius of a RGB star at the same luminosity (i.e., core mass, \citealt{bressan_parsec_2012}).
Our default model has a $\alpha_{\rm{MLT}}$ of 1.89. When we run a 1.0 $M_\odot$ model with $\alpha_{\rm{MLT}}$ of 1.7,
the radius versus core mass increases by up to 9 per cent, while using $\alpha_{\rm{MLT}}=2.0$ decreases the radius versus core mass by up to 5 per cent. 
\cite{tayar_correlation_2017} finds that $\alpha_{\rm{MLT}}$ ranges from about 1.74 to 2.06 for [Fe/H] ranging from -1 to 1, motivating the range we checked.
These changes in radius would affect $E_{\rm{bind}}$ and $\alpha_{\rm CE}$ by similar factors, but these remain much smaller than the uncertainty due to the unknown progenitor mass.

Convective overshoot on the main sequence (not included in our models) can slightly increase the core mass, causing stars to behave like slightly more massive stars on the RGB. Increasing overshoot would cause the lines in Figure \ref{fig:progenitors} to move to the left, slightly decreasing the progenitor masses and binding energies for the same core masses. However, this uncertainty is also dwarfed by the large possible range in progenitor mass.

\section{Conclusion}

For a sample of nine double WD binaries and one WD-BD binary, we model the primary WD's cooling age and thus constrain the system's ``birth period'' following common envelope (CE) evolution. Each system we analyze has a low-mass primary WD ($M_1 \! \lesssim \! 0.45 \, M_\odot$) which is likely a He-core WD. By considering all possible red giant progenitor stars for each WD primary, we constrain the range of possible CE ejection efficiencies, $\alpha_{\rm CE}$. Because of 1) the eclipsing nature of eight of the binaries and resulting low uncertainties for radii, temperatures, and masses, and  2) the lack of magnetic braking (as opposed to WD-M dwarf binaries), the binaries we model are some of the most promising candidates for precisely constraining CE ejection efficiency. 

We estimated the cooling age of the WD primary via a grid of WD models. Unlike some prior analyses that use published WD cooling tracks, we perform a comprehensive analysis that accounts for uncertainties such as the WD's hydrogen envelope mass and the possibility of H-burning flashes. Our two separate grids of He- and CO-core WDs incorporated element diffusion and gravitational settling, and assumed solar metallicity. We assumed GW-induced inspiral alone to convert the cooling age to the binary's post-CE birth period. We found that:

\begin{itemize}
    \item Our models that are consistent with the primary's radius and $T_{\rm{eff}}$ have masses that generally overlap well with the quoted mass $M_1$ from discovery papers.
    \item While hydrogen shell flashes may occur in the WD primary, models with hydrogen flashes and mass transfer do \textit{not} set the upper/lower bounds on birth periods.
    \item All but one system has a primary that is consistent with a low-mass CO WD as well as a He WD. The birth periods assuming a CO WD are larger than when assuming a He WD, due to the significant duration of the post-CE sdB phase preceding the CO WD's formation.
\end{itemize}

We created a mapping from our WD models to possible RGB progenitor stars, calculating the envelope binding energy when a viable progenitor has core mass equal to the core mass of the WD. We assumed the binding energy to be the sum of gravitational and internal (thermal plus recombination) energies, and the pre-CE period is calculated from the Roche-lobe overflow criterion. From the ratio of binding energy to the change in orbital energy (as mainly determined by the WD binary's birth period), we calculated $\alpha_{\rm CE}$. Our results are:

\begin{itemize}
    \item Assuming He WDs from RGB progenitors between 0.9 and 2.0 $M_\odot$ leads to a best-fitting constant $\alpha_{\rm CE}$ in the range $\sim$ 0.2 - 0.4. Therefore, our ten systems are consistent with a value of $\alpha_{\rm CE}$ that is substantially less than unity.
    
    \item The main uncertainty in $\alpha_{\rm CE}$ arises from the unknown progenitor star mass, followed by observational uncertainty in the primary WD mass (and hence progenitor core mass and envelope binding energy). Uncertainties stemming from the WD's hydrogen envelope mass, companion mass, metallicity, recombination energy, core boundary definition, modeling uncertainties, and other measurement uncertainties contribute at a lower level.
    
    \item Progenitors with $M \gtrsim 2.1 \, M_\odot$ require larger values of $\alpha_{\rm CE}$ because of the larger progenitor mass and smaller progenitor radius (hence much larger binding energies) when the CE event occurs. However, several systems have associated values of $\alpha_{\rm CE}$ significantly less than unity regardless of our inclusion of massive progenitors because massive progenitors are more limited in the masses of He WDs they can create. Therefore, considering the ensemble of systems, $\alpha_{\rm CE} < 0.5$  is still favored. 
    
    
    \item CO WD primaries from RGB progenitors between 1.8 and 2.2 $M_\odot$ (creating WDs of $\gtrsim$ 0.4 $M_\odot$) are relevant only to 4 systems. In this case, we find a best-fitting $\alpha_{\rm CE} \! \sim\! 0.5$. Progenitors of $M \gtrsim 2.3 \, M_\odot$ (which create CO WDs of $\gtrsim$ 0.32 $M_\odot$ and are relevant to nine systems) have associated $\alpha_{\rm CE}$ values larger than unity and are hence unlikely.

\end{itemize}

Our relatively low $\alpha_{\rm CE}$ values ($\alpha_{\rm CE}$ $\sim$ 0.2 - 0.4 for our main grid of He WDs) appear to be consistent with constraints from similar previous modeling of WD-WD, WD-M dwarf, and WD brown dwarf binaries. This motivates further investigation of whether a low $\alpha_{\rm CE}$ is consistent with other types of post-CE binaries, and the corresponding impact on predictions of binary population synthesis. Our work also encourages additional efforts to find and characterize short-period WD binaries, which are extremely useful for constraining $\alpha_{\rm CE}$. If a universal value of $\alpha_{\rm CE}$ is found to exist, a sample of short-period WD binaries with well constrained masses and temperatures will likely be a major cornerstone in supporting it.

\section*{Acknowledgements}

We are grateful for support from the NSF through grant AST-2205974. JF is thankful for support through an Innovator Grant from The Rose Hills Foundation, and the Sloan Foundation through grant FG-2018-10515. We thank Sterl Phinney and Kevin Burdge for useful discussions. We thank
the anonymous referee for their useful feedback and suggestions.

\section*{Data Availability}

 MESA inlists used in this work are provided in a Zenodo repository \citep{peter_scherbak_2022_7272761}. MESA models are available upon request to the corresponding author.

\label{conclusion}






\bibliographystyle{mnras}
\bibliography{main}

\appendix

\section{Effect of progenitor mass on  WD models}

\subsection{He models}
\label{He progen}

\begin{center}
\begin{table*}
\begin{tabular}{| c || c | c | c | c| c| c|} 
 \hline
 System &  0.9 \(M_\odot\) & 1.2  \(M_\odot\) & 2.0 \(M_\odot\)& 2.3 \(M_\odot\)& 2.5 \(M_\odot\)& 2.8 \(M_\odot\)\\

 \hline
  ZTF J2029 & 27.9 - 36.7  & 27.2 - 37.6   & 31.6 - 41.3 & 44.9 - 54.2  & 47.1 - 63.2 & 51.8 - 59.8 \\
 \hline
 SDSS J0822 & 75.8 - 83.4  & 75.1 - 83.8  & 78.2 - 86.8 & 85.4 - 94.2 & 86.7 - 97.0 & 89.7 - 98.4 \\

 \hline

\end{tabular}

 \caption{The inferred range of post-CE birth periods (in minutes) for two systems, using progenitors of different initial masses. Top row refers to the initial mass of the progenitor used to create a grid of 0.3 $M_\odot$ He WD models. }
\label{tab:He progen .3}
\end{table*}

\end{center}







To characterize the effect of progenitor mass on the behavior of WD models, we created a grid of 0.3 $M_\odot$ He models from several different progenitor masses: 0.9, 1.2, 2.0, 2.3, 2.5 and 2.8 $M_\odot$. We found that models created from higher mass RGB stars tended to cool slower, as expected from considerations that the core is less degenerate at mass stripping and contracts further following the WD's formation, releasing thermal energy. For models from 0.9 - 2.0 $M_\odot$ progenitors, the effect of progenitor mass on the WD cooling and inferred birth period was small compared to overall uncertainties.

However, using models from a 2.2 $M_\odot$ progenitor started to increase the inferred birth period substantially, and using more massive progenitors continued the increase (Table \ref{tab:He progen .3}). We use ZTF J2029 and SDSS J0822 as benchmarks because they are well-matched by 0.3 $M_\odot$ He models.
Additionally, the mass of H that formed the boundary between flashing and non-flashing models remained similar for 0.9 - 2.0 $M_\odot$-generated models, but began to change once a progenitor mass of 2.2 $M_\odot$ was reached.

When calculating values of $\alpha_{\rm{CE}}$, we use the grid of WD models generated from a 1.2 $M_\odot$ progenitor. This is justified because the dominant effect of a massive ($>2 M_\odot$) progenitor will be to increase $E_{\rm{bind}}$ and therefore increase $\alpha_{\rm{CE}}$, with the change in WD models/cooling ages a secondary effect.  In the case of ZTF J2029, for example, the effect on $\alpha_{\rm CE}$ from different WD cooling  models (generated with a 2.0  $M_\odot$  vs 2.3 $M_\odot$  progenitor)  will be less than about 30 per cent due to the difference in associated birth period. The change in $E_{\rm{bind}}$ when comparing a 2.0  $M_\odot$  vs 2.3 $M_\odot$ progenitor is generally much larger and more important. Therefore, we use the same grid of WD models to model the WD primary (i.e. finding the same cooling age/birth period) and to perform the calculation of $\alpha_{\rm CE}$ even when considering  high-mass progenitors.

\subsection{CO models}

\label{CO progen append}

We compared 0.4/0.45 $M_\odot$ CO WD models made from a 3.0 $M_\odot$ progenitor (in the blue CE-forbidden region of Fig. \ref{fig:progenitors}) to models made from 3.5/4.0 $M_\odot$ progenitors (in the pink CE-allowed region). The differences occur predominantly in the duration of the sdB phase (which makes sense, given the choice of progenitor determines whether He-burning has already begun before mass stripping). However, the changes  are small (a few times $10^{7}$ years) compared to the overall duration of the sdB phase (several  times $10^{8}$ years) and do not greatly affect our results.
Therefore, even though we do not expect a 3.0 $M_\odot$ progenitor to create a 0.4 or 0.45 $M_\odot$ CO WD through CE evolution (as those are core masses past the tip of the RGB but preceding the AGB), it is reasonable for our purposes to use a 3.0 $M_\odot$ progenitor model.

Similarly, we compared 3.0 $M_\odot$ progenitor models with sdB models formed near the tip of the RGB from 1.0 and 2.0 $M_\odot$ progenitors. It was difficult to run stripped star models that flash \textit{after} mass stripping. While we could create such models, and track the He flash (therefore identifying them as about to become sdB stars), numerical difficulties prevented us from running them through the duration of the sdB. However, as in the MESA test suite and also \cite{bauer_phases_2021}, we could create and run models that flash just \textit{before} mass stripping.

\begin{figure}
    \centering
    \includegraphics[scale=.5]{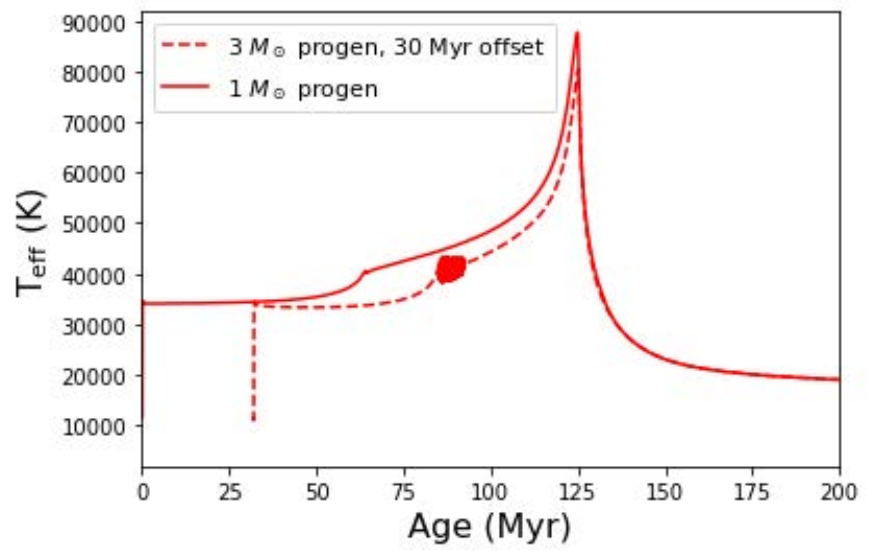}
    \caption{Two 0.47 $M_\odot$ CO models, one from a 1 $M_\odot$ progenitor stripped just after He ignition (close to canonical) and one from 3 $M_\odot$ progenitor. The second model shows oscillations due to He-burning flashes (discussed in \ref{sdB}). The duration of the sdB phase changes (by the offset of 3e7 years), but the WD cooling curves are not significantly affected by the choice of progenitor.}
    \label{fig:co_progen}
\end{figure}

Therefore, we compared 3.0 $M_\odot$ progenitor, 0.44/0.47 $M_\odot$ WD models (in the blue CE-forbidden region) to 2.0 $M_\odot$ progenitor, $\approx$ 0.44 $M_\odot$ WD models and 1.0 $M_\odot$ progenitor, $\approx$ 0.47 $M_\odot$ WD models (in both cases stripped slightly after the tip of the RGB). Models from a 2.0 $M_\odot$ progenitor were similar to models from a 3.0 $M_\odot$ progenitor. Models from a 1.0 $M_\odot$ progenitor showed a significantly longer sdB phase versus the 3.0 $M_\odot$ progenitor models because the 3.0 $M_\odot$ progenitor models had already burnt much of their He at the time of mass stripping. However, the post-sdB behavior (i.e. the cooling CO WD) showed similar behavior (e.g. Fig. \ref{fig:co_progen}). Therefore, we would expect 0.47 $M_\odot$ models from two different progenitors to either be both good fits or both bad fits to the observational constraints. Therefore, as no 0.47 $M_\odot$ WD models with a 3.0 $M_\odot$ progenitor were matches to our systems, they would likely not be matches even if created with a  1.0 $M_\odot$ progenitor, and we did not create a full grid of canonical sdB models.

\section{Creation of He WD models}

Our He WD models are created through mass stripping of an evolved RGB star. We do not assume \textit{ab initio} the maximum amount of H expected to be found on a cooling WD. In the case of a large amount of H on the model, the model will remain radially extended and its surface temperature will remain elevated (i.e. the model does not cool) until excess H burns. One method is to let the model adjust in isolation, burning any excess H, then save the model immediately after the surface temperature began to decrease, so that further evolution in the binary starts with the WD primary at the top of its cooling track. The other method is to place the  model into a binary as it adjusts, where we find that excess H is lost to mass transfer to the companion on a time-scale of hundreds of years, after which the model begins to cool/contract. Comparing the two methods
showed that they result in similar cooling behavior for the model after a time much shorter than the ages we infer, so this initial transient phase is unimportant for our purposes. 

For our grid of models, we choose to model them in isolation,
saving them for further use when their temperatures begin to decrease. However, we still need a criterion so that the adjustment phase does not extend arbitrarily long for large amounts of H. Therefore, we reject models that expand beyond 1 $R_{\odot}$ as they will almost certainly undergo mass transfer and lose the excess H to the companion. This limits the adjustment phase to less than 1 Myr and makes it irrelevant to estimating cooling ages. However, it is still important to model this phase to see how much H can survive, which influences the WD cooling behavior.

\subsubsection{Creation of models with low masses of H}

To create models with a very low mass of H, we take a relatively low-H model as created above in Sec. \ref{create_He} (e.g. a 0.3198 $M_\odot$ core, 0.32 $M_\odot$ model) and replace most of its outer H envelope with He, via \textit{replace\_element = .true.} and setting the boundaries \textit{replace\_element\_nzhi} and \textit{replace\_element\_nzlo} to target the inner regions of the H envelope. This method was used to create models with masses of H below around $10^{-6}$ $M_\odot$.

\section{Effect of metallicity on He WD models}

\label{metallicityHe}

\begin{figure}
    \centering
    \includegraphics[scale=.2]{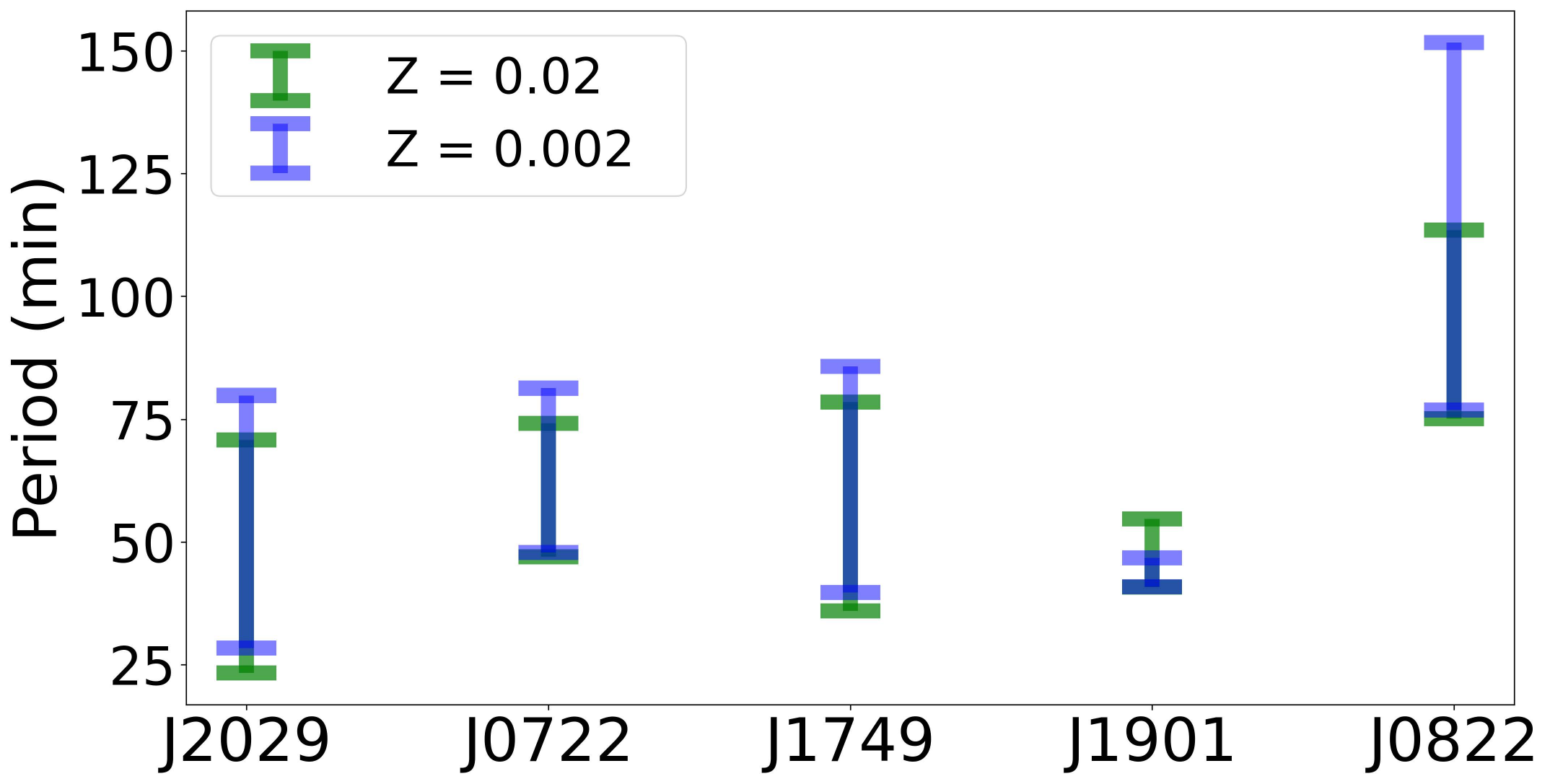}
    \caption{A lower metallicity grid results in longer possible birth periods for most of the systems that we model. Not all systems are shown because some have low birth period uncertainties when modeled at either metallicity.  }
    \label{fig:compz}
\end{figure}

We have assumed solar metallicity (Z=0.02) in the creation of all WD models. However, a lower metallicity can change WD cooling behavior. At lower metallicity, the threshold of H required to undergo a H flash increases - i.e. there are models that flash at Z=0.02, as opposed to those that maintain stable burning at Z=0.002.  Therefore, the main effect of lowering metallicity is to introduce slowly-cooling models with high H, that would have flashed and then cooled more quickly at higher metallicity. This leads to higher possible birth periods for several of the systems that we model (Fig. \ref{fig:compz}). We show only half the systems we model for clarity, because the others have relatively small period uncertainties for either metallicity grid. The change is largest in the case of SDSS J0822 -  models with a large H envelope can take a very long time to cool to its relatively low $T_{\rm{eff,1}} \simeq\,$14,000 K. Interestingly, this is the only binary we model where the discovery paper claims it to be located in the galaxy's halo \citep{brown_discovery_2017}. Therefore, it is possible that it does indeed have low metallicity, and the upper period quoted for our main grid of solar-metallicity models is an underestimate. The effect of this change, combined with the change in envelope binding energy for progenitor stars of lower metallicity, is discussed in Sec. \ref{progenitor discuss}.

\section{Flashing He WD models: Extra Details}

\label{appendix flashes}

Flashing WDs undergo a large expansion in radius that will likely lead to mass transfer with the companion. When modeled, mass transfer was computed via \textit{mdot\_scheme = Kolb}. We compared the Kolb and Ritter mass loss schemes for a single model, and found the mass loss and resulting cooling behavior nearly identical, but Ritter requiring more timesteps and computing time. Additionally, we used MESA release 10398 instead of more recent versions, because of its empirical increased computational speed for flashes and mass transfer.


We tested the effect of changing the orbital period and the companion mass on our simulations. Performing a subset of simulations with $M_2$ of 0.57 $M_\odot$ (the highest feasible $M_2$ from Table \ref{tab:binary parameters}) made negligible difference with regard to mass loss/post-flash cooling curves.  We also ran a subset of models at (more computationally difficult)  periods of 60-70 minutes, and found that the amount of H lost or the post-flash cooling behavior was not greatly changed when compared to models at 100 minute periods. Because the flashing model's envelope wants to expand so greatly beyond the RL (i.e. it would expand to 10s of $R_\odot$ if not modelled in a binary), the exact value of the RL does not matter greatly. As expected, we found that any of these additional models which matched to our systems could be bounded above and below in cooling age/birth period by non-flashing models.

The one exception is ZTF J1901, where flashing models (fitting to observed radius and $T_{\rm{eff}}$) \textit{do} provide the upper cooling age bound for the system.  However, in these cases the upper-bound flashing models only match the observed temperature during the spike in temperature immediately after the flash (such a spike is demonstrated in the nearly vertical red lines in Fig. \ref{fig:flash_vs_noflash}). The time-scale for this spike in temperature (which is similar to the time-scale of the large loop in the HR diagram in Fig. \ref{fig:flash HR}) is less than $10^5$ years and these models match the observed temperature for only about $10^4$ years. In contrast, the length of time non-flashing models match temperature constraints is closer to $10^6$ years for this system. Therefore, it is far less likely that we are observing the system in an immediately post-flash state, as compared to undergoing normal cooling behavior. We reject such flashing models, which show similarly small ages of relevance, for \textit{all} systems we  model.

\section{Effect of He core boundary definition on envelope binding energy}

\label{Core boundary effect}

\begin{figure}
     \centering
     \includegraphics[scale=0.6]{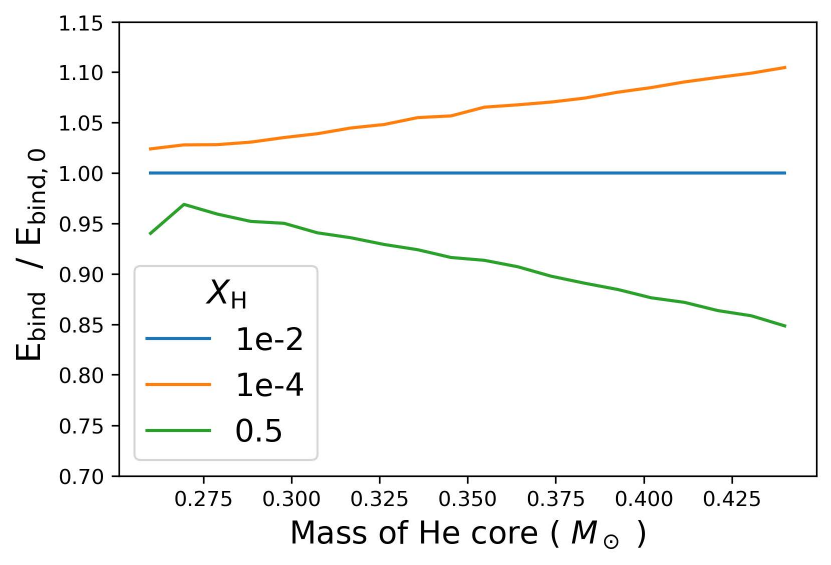}
     \caption{Effect of core/envelope boundary definition on envelope binding energy. $X_{\rm H}$ refers to the mass fraction of H below which the He core is defined to begin. $E_{\rm bind, 0}$ refers to envelope binding energy with our default definition for core/envelope boundary ($X_{\rm H} = 10^{-2}$).}  
     \label{fig:ebind_definitions}
\end{figure}

An additional uncertainty associated with calculating the binding energy of the H envelope $E_{\rm{bind}}$ is the exact definition of the envelope/core boundary. For both WD and RGB models, we have used MESA's default definition of the He core to be where the mass fraction of H, $X_{\rm{H}}$, drops below 0.01. The effect of changing this definition on $E_{\rm{bind}}$ is shown in Fig. \ref{fig:ebind_definitions}. To calculate this, we iterate through the saved profiles for a 1.5 $M_\odot$ RGB star, finding when the core reaches a certain mass by a certain definition (e.g. when the core reaches 0.4 $M_\odot$ for $X_{\rm{H}}=0.5$). As before, we then integrate down to that mass coordinate to find $E_{\rm{bind}}$. Using a smaller value of $X_{\rm{H}}$ decreases the core mass for a given model and therefore increases $E_{\rm{bind}}$, and vice versa. The change depends on core mass, but can reach up to 20 per cent. However, this uncertainty is small compared to the uncertainty from the unknown total mass of the RGB star.





\bsp	
\label{lastpage}
\end{document}